\documentclass[12pt]{amsart}
\usepackage{geometry}
\usepackage{amsaddr}

\usepackage{amsmath}
\usepackage{amssymb}

\usepackage{graphicx}
\usepackage[utf8]{inputenc}     
\usepackage[T1]{fontenc}        
\allowdisplaybreaks             

\usepackage[numbers,sort&compress]{natbib}

\usepackage{tikz}
\usetikzlibrary{quantikz2}      
\newcommand{\DAH}{disable auto height}
\newcommand{\myquantikz}[1]{
\begin{tikzpicture} \node[scale=0.78] { \begin{quantikz}[row sep={0.8cm,between origins}]
#1
\end{quantikz} }; \end{tikzpicture}
}

\usepackage{hyperref}
\hypersetup{
 unicode=true,
 pdftoolbar=false, pdfmenubar=true, pdffitwindow=false, pdfstartview={FitH},
 pdftitle={Local qubit invariants on quantum computer},
 pdfauthor={Frédéric Holweck, Szilárd Szalay},
 pdfsubject={research article on local qubit invariants and covariants},
 pdfkeywords={multipartite entanglement} {local unitary invariants} {experiment on quantum computer} {IBM},
 pdfcreator={pdflatex}, pdfproducer={overleaf},
 pdfnewwindow=true, colorlinks=true, linktoc=page, linkcolor=blue, citecolor=blue, filecolor=blue, urlcolor=blue
}

\usepackage{braket}

\newcommand{\nn}[1]{\bar{#1}}

\DeclareMathOperator{\Tr}{Tr}
\DeclareMathOperator{\Det}{Det}

\newcommand{\field}[1]{\mathbb{#1}}

\newcommand{\norm}[1]{\Vert #1 \Vert}

\newcommand{\abs}[1]{\vert #1 \vert}

\newcommand{\Bigabs}[1]{\Bigl\vert #1 \Bigr\vert}

\newcommand{\T}{\mathrm{T}}

\begin{document}

\title{Local qubit invariants on quantum computer}
\author{Szilárd Szalay}
\email{szalay.szilard@wigner.hu}
\address{Department of Theoretical Solid State Physics, Wigner Research Centre for Physics, Budapest, Hungary}
\author{Frédéric Holweck}
\email{frederic.holweck@utbm.fr}
\address{Laboratoire Interdisciplinaire Carnot de Bourgogne,  UMR 6303 CNRS, University of Technology of Belfort-Montbéliard, 90010 Belfort Cedex, France}
\address{Mathematics and Statistics Department, Auburn University, Auburn, AL, USA}

\begin{abstract}
We present two general methods to implement quantum circuits for the direct measuring of local unitary invariants on quantum computers.
We work these out for important three-qubit invariants,
and also demonstrate these on the IBM Quantum Platform for important entanglement measures of three qubits.
\end{abstract}

\maketitle

\section{Introduction}
\label{sec:intro}

Entanglement is the most notable form of nonclassical correlations characteristic of quantum systems~\cite{Horodecki-2009}.
Since entanglement is invariant with respect to local unitary transformations,
local unitary (LU-) invariants give the finest description of entanglement, in principle,
everything about entanglement in a quantum state can be told in terms of LU-invariants.
A significant amount of effort was therefore made to apply invariant theory to multipartite quantum systems~\cite{Linden-1998,Grassl-1998,Luque-2007,Williamson-2011,Osterloh-2010,Eltschka-2012,Osterloh-2012},
important exotic results were obtained for small systems~\cite{Carteret-1999,Kempe-1999,Coffman-2000,Sudbery-2001,Luque-2003,gour2010all, chen2013proof,Szalay-2012,Szalay-2025},
while also the general description was worked out~\cite{Hero-2009,Hero-2011,Vrana-2011,Vrana-2011b,johansson2014local,oeding2025tensor, carrozza2026tensor}.

Recently noisy intermediate scale quantum (NISQ) computers became accessible online,
and entangled quantum systems can be manipulated now from home,
making possible the demonstration of several exotic quantum phenomena~\cite{alsina2016experimental,swain2019experimental,devitt2016performing,holweck2021testing,jusseau2025four,kelleher2025empirical}. 
In this work we describe \emph{two general methods} for the implementation
of the \emph{direct measurement} of important nonnegative real-valued LU-invariants of $n=2$ and $3$ qubits on quantum computers, and we also demonstrate these using the IBM Quantum Platform~\cite{IBMQuantum-2026,qiskit2024}.
These LU-invariants are positively homogeneous functions of degree $k=4,8,12$,
and the appropriate roots of some of these are important \emph{entanglement measures}~\cite{Wootters-1998,Coffman-2000,Dur-2000,Szalay-2025}.
The two methods use $nk/4$ and $nk/2$ qubits, respectively, and both use $k/2$ copies of the state.
More precisely, as cloning quantum states is impossible,
these use $k/2$ instances of the oracle $U_\psi$ producing the state and/or its transpose or adjoint or conjugate.
When we use the word ``copy'' of the state, we always mean multiple instances of the oracles, preparing the states.
The different instances may produce states with different error, we provide error estimates in Appendix~\ref{sec:error}.
Note that the total computational power needed for the circuits is not known in general, as it depends on the concrete implementation of the oracle $U_\psi$, depending on the concrete $\ket{\psi}$ to be prepared.
The second method leads to larger circuits, utilizing a larger number of CNOT gates, and measurements on twice as many qubits,
leading to lower precision, or higher relative error on the outcomes than the first one.

The methods presented in this work translate the \emph{index contractions} in the definitions of the LU-invariants
to the language of quantum circuits,
then the probability of a particular outcome of the measurement of the qubit registers gives a power of the value of the LU-invariant directly.
Our first method is related to the \emph{interferometric method}~\cite{Horodecki-2002,Ekert-2002,Leifer-2004},
where the value of the LU invariant can be read off from the visibility of the interference pattern (by maximization/fitting).
Other methods on the evaluation of LU-invariants on quantum computers were given by \emph{ad-hoc methods}
based on the tracking of the terms of the state vector coefficients 
in the polynomials by which the LU-invariants are formulated.
These were given for example for the concurrence squared~\cite{Romero-2007,Karimi-2024} in two-qubit systems,
or for the three-tangle squared~\cite{Yahyavi-2022,Elyasi-2025} in three-qubit systems.
The resulting circuits are related to our second method.
Besides these \emph{direct methods}, there is another method, relying on LU-canonical forms, which involves \emph{optimization}~\cite{Perez-Salinas-2020},
which means a sufficiently large number of runs of the protocol for different parameters
to achieve the optimized value.
There is also another way,
LU-invariants can of course be evaluated simply by substituting the amplitudes resulted from the \emph{full tomography} of the state~\cite{Banaszek-2013}.
We emphasize that the contraction methods presented here are applicable to arbitrary number of subsystems of arbitrary dimensions,
and are especially useful 
when the ad-hoc, optimization based, and tomography based methods become infeasible with the increase of the system size.

\section{Local invariants and covariants}
\label{sec:theory}

We consider polynomial invariants and covariants of multiqubit quantum states $\ket{\psi}$
with respect to the \emph{local} transformations $\ket{\psi}\mapsto \ket{\psi'}=G_1\otimes G_2\otimes\dots\otimes G_n\ket{\psi}$,
where the operators $G_a$ are unitary ($G_a\in\mathrm{U}(2)$) or of determinant $1$ ($G_a\in\mathrm{SL}(2)$).
Polynomial $\mathrm{U}(2)$ and $\mathrm{SL}(2)$ invariants (scalars) and covariants (tensors) of qubit systems can be formulated by index contractions of the state vectors with
\begin{subequations}
\begin{align}
    \label{eq:delta}
    \delta   &= \begin{bmatrix} 1 & 0 \\ 0 & 1 \end{bmatrix}   = I
\intertext{and}
    \label{eq:epsilon}
    \epsilon &= \begin{bmatrix} 0 & 1 \\ -1 & 0 \end{bmatrix}  = i Y,
\end{align}
\end{subequations}
since for all $2\times 2$ unitary matrix $U$,
we have $U^\dagger \delta U = \delta$,
and for all $2\times 2$ matrix $M$,
we have $M^\T \epsilon M = \det(M)\epsilon$.
Here we show general methods, translating the index contractions directly to the language of quantum circuits.

\subsection{Two qubits}
\label{sec:theory.22}

Let us begin with the case of two qubits, to show the main idea and the tools used.
For two-qubit state vectors $\ket{\psi} = \sum_{i,j=0}^1\psi^{ij}\ket{ij}$ we have two important invariants,
the norm-squared and (two times of) the determinant,
\begin{subequations}
\label{eq:22tens}
\begin{align}
\label{eq:22tens.n}
    n^2(\psi) := \norm{\psi}^2
    = \sum_{ii'jj'} \delta_{ii'}\delta_{jj'}\overline{\psi^{ij}}\psi^{i'j'}
    &= \overline{\psi^{00}}\psi^{00}+\overline{\psi^{01}}\psi^{01}+\overline{\psi^{10}}\psi^{10}+\overline{\psi^{11}}\psi^{11},\\
    \label{eq:22tens.q}
    q(\psi) := 2\det(\psi)
    = \sum_{ii'jj'} \epsilon_{ii'}\epsilon_{jj'}\psi^{ij}\psi^{i'j'}
    &= \psi^{00}\psi^{11}-\psi^{01}\psi^{10}-\psi^{10}\psi^{01}+\psi^{11}\psi^{00},
\end{align}
\end{subequations}
being $\mathrm{U}(4)$ and $\mathrm{SL}(2)\times\mathrm{SL}(2)$ invariants, respectively.
On the other hand, both are invariant with respect to the permutation of the subsystems.

We will consider the absolute value square of the invariants~\eqref{eq:22tens},
\begin{subequations}
\label{eq:22invs}
\begin{align}
    \label{eq:22invs.n}
    n^4(\psi) &:= \norm{\psi}^4,\\
    \label{eq:22invs.c}
    c^2(\psi) &:= 4\abs{\det(\psi)}^2,
\end{align}
\end{subequations}
as these are the ones being measurable by our methods.
The \emph{concurrence squared} $c^2(\psi)$ is of direct importance in entanglement theory~\cite{Wootters-1998,Szalay-2013},
being an entanglement measure, 
the linear entropy $2(\Tr(\rho_1)^2-\Tr(\rho_1^2))=4\det(\rho_1)=4\abs{\det(\psi)}^2$
of the reduced state $\rho_1:=\Tr_2(\ket{\psi}\bra{\psi})$.
The concurrence $c(\psi)=2\abs{\det(\psi)}$ itself is also an entanglement measure, 
being an increasing concave function of the usual entanglement entropy~\cite{Bennett-1996} for qubits~\cite{Wootters-1998},
gained relevance recently~\cite{Szalay-2025}.
The normalization is often relaxed in invariant theory,
while we have $n(\psi) = 1$ in quantum probability theory.
Note, however, that $n(\psi)<1$ may occur even in the latter case,
if it is obtained by a quantum circuit, plagued by errors.
Then $n(\psi)$ can be used to estimate the error during the run.

\begin{figure}
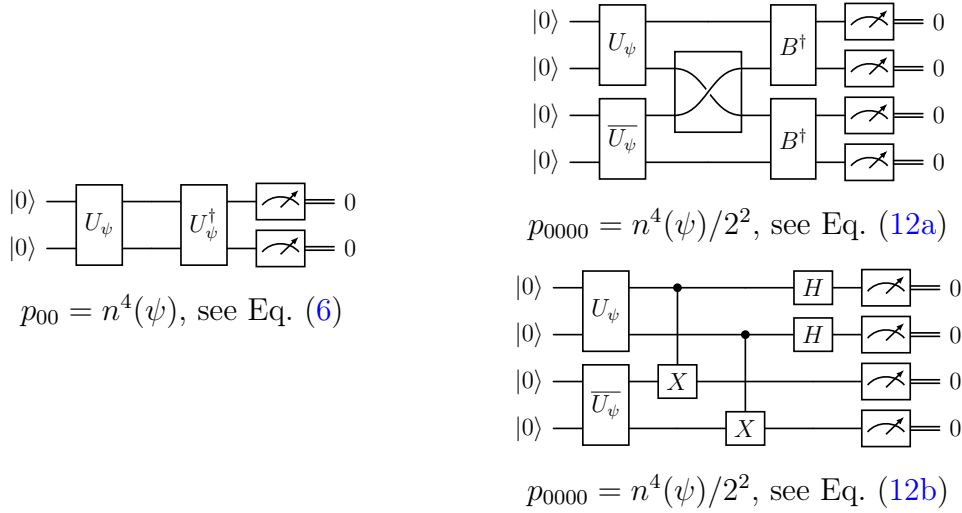

\begin{minipage}{0.48\textwidth}
\centering
    \myquantikz{
    \lstick{$\ket{0}$} & \gate[2,\DAH]{U_\psi} & & \gate[2,\DAH]{U_\psi^\dagger}
     & \meter{} & \setwiretype{c} \rstick{$0$}  \\
    \lstick{$\ket{0}$} &    &    &
     & \meter{} & \setwiretype{c} \rstick{$0$}
    }\\
    $p_{00} = n^4(\psi)$, see Eq.~\eqref{eq:p22.n} 
\end{minipage}
\begin{minipage}{0.48\textwidth}
\centering
    \myquantikz{
    \lstick{$\ket{0}$} & \gate[2,\DAH]{U_\psi} &    & \gate[2,\DAH]{B^\dagger}
     & \meter{} & \setwiretype{c} \rstick{$0$}  \\
    \lstick{$\ket{0}$} &    & \gate[2,swap]{} &
     & \meter{} & \setwiretype{c} \rstick{$0$}  \\
    \lstick{$\ket{0}$} & \gate[2,\DAH]{\overline{U_\psi}} &    & \gate[2,\DAH]{B^\dagger}
     & \meter{} & \setwiretype{c} \rstick{$0$}  \\
    \lstick{$\ket{0}$} &    &    &
     & \meter{} & \setwiretype{c} \rstick{$0$}
    } \\
    $p_{0000} = n^4(\psi)/2^2$, see Eq.~\eqref{eq:p22B.n0}
    \myquantikz{
    \lstick{$\ket{0}$} & \gate[2,\DAH]{U_\psi} & \ctrl{2} &  & \gate{H}
     & \meter{} & \setwiretype{c} \rstick{$0$}  \\
    \lstick{$\ket{0}$} &                       &  & \ctrl{2} & \gate{H}
     & \meter{} & \setwiretype{c} \rstick{$0$}  \\
    \lstick{$\ket{0}$} & \gate[2,\DAH]{\overline{U_\psi}} & \gate{X} &  & 
     & \meter{} & \setwiretype{c} \rstick{$0$}  \\
    \lstick{$\ket{0}$} &                       &  & \gate{X} &
     & \meter{} & \setwiretype{c} \rstick{$0$}
    } \\
    $p_{0000} = n^4(\psi)/2^2$, see Eq.~\eqref{eq:p22B.n}
\end{minipage}
\caption{Quantum circuits implementing the two-qubit norm-squared \eqref{eq:22invs.n}.}
\label{fig:22.n}
\end{figure}

First we need a $\mathrm{U}(4)$ unitary gate, performing the preparation of the state $\ket{\psi}$.
In particular, 
\begin{equation}
\label{eq:Upsi22}
    U_\psi := \sum_{i,j = 0}^1\ket{\chi_{ij}}\bra{ij},
\end{equation}
where $\ket{\chi_{ij}}$ are orthonormal, $\braket{\chi_{ij}|\chi_{i'j'}} = \delta_{ii'}\delta_{jj'}$,
with $\ket{\chi_{00}} := \ket{\psi}$ fixed to the state to be prepared.
Then $U_\psi$ acting on the registers initialized to $\ket{00}$ leads to the desired state, $U_\psi\ket{00} = \ket{\chi_{00}} = \ket{\psi}$.
To obtain the norm-squared $n^2(\psi)$, we need to act on this with $\bra{\psi} = \bra{00}U_\psi^\dagger$,
which comes by acting with
\begin{equation}
\label{eq:Upsi22Adj}
    U_\psi^\dagger := \sum_{i,j = 0}^1\ket{ij}\bra{\chi_{ij}},
\end{equation}
then the $00$ component of the resulting vector $U_\psi^\dagger U_\psi\ket{00} = n^2(\psi)\ket{00}$ is the norm-squared $n^2(\psi)$.
Then performing measurements on each qubit in the computational basis,
the $p_{00}$ probability of the outcome $00$ is just the square of the norm squared,
\begin{equation}
\label{eq:p22.n}
    p_{00} = \abs{\bra{00}U_\psi^\dagger U_\psi\ket{00}}^2 = n^4(\psi).
\end{equation}
(The quantum circuit implementing this method is shown in the left part of Figure~\ref{fig:22.n}.)
Note again that this can be less than $1$ due to errors, then also $p_{ij}>0$ for at least one outcome $ij\neq00$.

To have the concurrence-squared~\eqref{eq:22invs.c}, 
we exploit the formula~\eqref{eq:22tens.q} given by local index contractions with $\epsilon$,
which is just the Pauli-Y operator up to a phase \eqref{eq:epsilon}.
After preparing the state as before, $U_\psi\ket{00} = \ket{\psi}$,
we need to act on this with Pauli-Ys, $Y\otimes Y\ket{\psi} = -\sum_{ii'jj'}\epsilon_{ii'}\epsilon_{jj'}\psi^{i'j'}\ket{ij}$,
and contract the indices with those of the original state vector.
Note that, contrary to the case of the norm, applying $U_\psi^\dagger$ is not suitable here,
since it contains $\bra{\psi} = \sum_{i,j = 0}^1\overline{\psi^{ij}}\bra{ij}$
with the $\overline{\psi^{ij}}$ complex conjugates of the amplitudes $\psi^{ij}$.
We have two methods to use here.

\begin{figure}
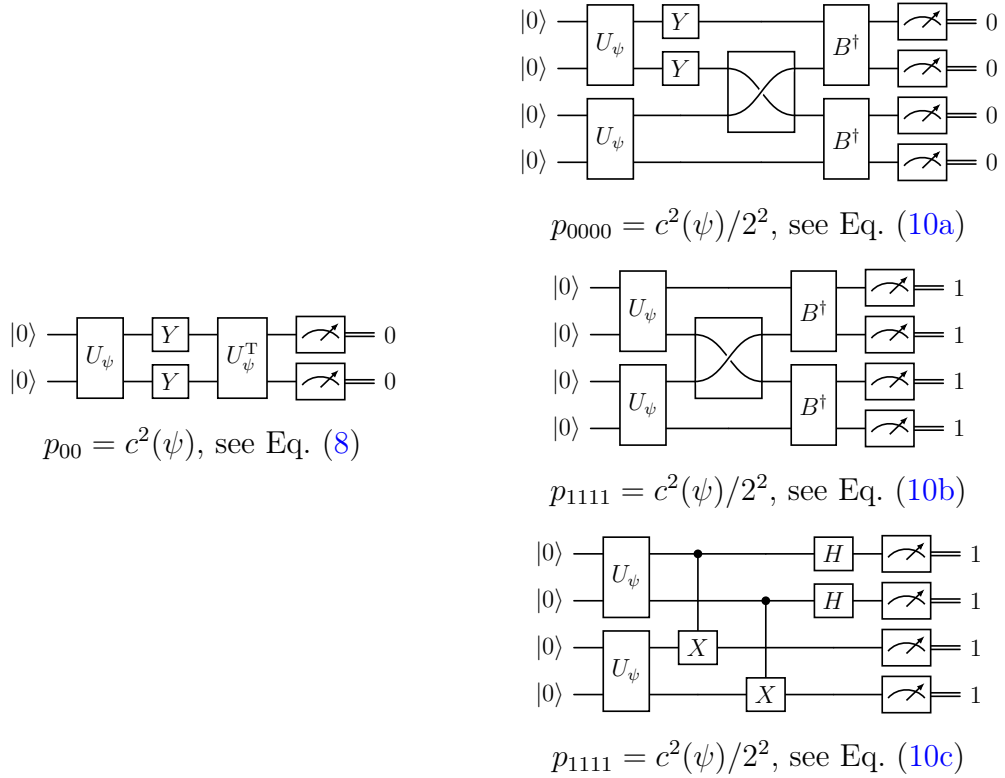

\begin{minipage}{0.48\textwidth}
\centering
\myquantikz{
    \lstick{$\ket{0}$} & \gate[2,\DAH]{U_\psi} & \gate{Y} & \gate[2,\DAH]{U_\psi^\T}
     & \meter{} & \setwiretype{c} \rstick{$0$}  \\
    \lstick{$\ket{0}$} &                  & \gate{Y} &
     & \meter{} & \setwiretype{c} \rstick{$0$}
    } \\
    $p_{00} = c^2(\psi)$, see Eq.~\eqref{eq:p22.c}
\end{minipage}
\begin{minipage}{0.48\textwidth}
\centering
    \myquantikz{
    \lstick{$\ket{0}$} & \gate[2,\DAH]{U_\psi} & \gate{Y} &  & \gate[2,\DAH]{B^\dagger}
     & \meter{} & \setwiretype{c} \rstick{$0$}  \\
    \lstick{$\ket{0}$} & & \gate{Y} & \gate[2,swap]{} &
     & \meter{} & \setwiretype{c} \rstick{$0$}  \\
    \lstick{$\ket{0}$} & \gate[2,\DAH]{U_\psi} &  &  & \gate[2,\DAH]{B^\dagger}
     & \meter{} & \setwiretype{c} \rstick{$0$}  \\
    \lstick{$\ket{0}$} & &   &    &
     & \meter{} & \setwiretype{c} \rstick{$0$}
    }\\
    $p_{0000} = c^2(\psi)/2^2$, see Eq.~\eqref{eq:p22B.c0}
    \myquantikz{
    \lstick{$\ket{0}$} & \gate[2,\DAH]{U_\psi} &    & \gate[2,\DAH]{B^\dagger}
     & \meter{} & \setwiretype{c} \rstick{$1$}  \\
    \lstick{$\ket{0}$} &    & \gate[2,swap]{} &
     & \meter{} & \setwiretype{c} \rstick{$1$}  \\
    \lstick{$\ket{0}$} & \gate[2,\DAH]{U_\psi} &    & \gate[2,\DAH]{B^\dagger}
     & \meter{} & \setwiretype{c} \rstick{$1$}  \\
    \lstick{$\ket{0}$} &    &    &
     & \meter{} & \setwiretype{c} \rstick{$1$}
    }\\
    $p_{1111} = c^2(\psi)/2^2$, see Eq.~\eqref{eq:p22B.c1}
        \myquantikz{
    \lstick{$\ket{0}$} & \gate[2,\DAH]{U_\psi} & \ctrl{2} &  & \gate{H}
     & \meter{} & \setwiretype{c} \rstick{$1$}  \\
    \lstick{$\ket{0}$} &                       &  & \ctrl{2} & \gate{H}
     & \meter{} & \setwiretype{c} \rstick{$1$}  \\
    \lstick{$\ket{0}$} & \gate[2,\DAH]{U_\psi} & \gate{X} &  & 
     & \meter{} & \setwiretype{c} \rstick{$1$}  \\
    \lstick{$\ket{0}$} &                       &  & \gate{X} &
     & \meter{} & \setwiretype{c} \rstick{$1$}
    }\\
    $p_{1111} = c^2(\psi)/2^2$, see Eq.~\eqref{eq:p22B.c}
\end{minipage}
\caption{Quantum circuits implementing the two-qubit concurrence \eqref{eq:22invs.c}.}
\label{fig:22.c}
\end{figure}

The \textit{first method} is to act with $\bra{\overline{\psi}} = \bra{00}U_\psi^\T$,
where $U_\psi^\T$ is the transposition of the unitary gate $U_\psi$,
\begin{equation}
\label{eq:Upsi22Transp}
    U_\psi^\T := \sum_{i,j = 0}^1\ket{ij}\bra{\overline{\chi_{ij}}},
\end{equation}
where the complex conjugation (and therefore the transposition) is understood in the computational basis, 
$\ket{\overline{\psi}} = \sum_{i,j = 0}^1\overline{\psi^{ij}}\ket{ij}$,
then
$\bra{\overline{\chi_{00}}} = \bra{\overline{\psi}} = \sum_{i,j = 0}^1\psi^{ij}\bra{ij}$,
just as it is needed.
So, after acting with $U_\psi^\T$ and 
performing measurements on each qubit in the computational basis,
the $p_{00}$ probability of the outcome $00$ is just the concurrence squared,
\begin{equation}
\label{eq:p22.c}
    p_{00} = \abs{\bra{00}U_\psi^\T (Y\otimes Y) U_\psi\ket{00}}^2 = c^2(\psi).
\end{equation}
(The quantum circuit implementing this method is shown in the left part of Figure~\ref{fig:22.c}.)

The \emph{second method} may be useful if implementing $U_\psi^\T$ cannot be done, or is infeasible,
for example, when analyzing the output of an involved protocol.
Here we need two more qubit registers storing another copy of the state $\ket{\psi}$
(another instance of $U_\psi$), then $Y\otimes Y$ acts on one of these,
$((Y\otimes Y)\ket{\psi})\otimes\ket{\psi} = (Y\otimes Y\otimes I\otimes I)(U_\psi\otimes U_\psi)\ket{0000}$,
then the contractions of the respective indices of the two copies are needed.
Index contraction can be done by acting with the bra $\bigl(\sum_{ij = 0}^1\delta_{ij}\bra{ij}\bigr) = \bra{00}+\bra{11} = \sqrt{2}\bra{\text{Bell}_0}$
on the respective qubits,
which can be achieved by the $B^\dagger$ adjoint of the Bell state preparing unitary
\begin{equation}
\label{eq:B}
    B = \ket{\text{Bell}_0}\bra{00} + \ket{\text{Bell}_1}\bra{01} + \ket{\text{Bell}_3}\bra{10} + i\ket{\text{Bell}_2}\bra{11}
    = C (H\otimes I),
\end{equation}
then selecting the $00$ outcome of the measurements in the computational basis selects $\bra{\text{Bell}_0}$ for the index contraction.%
\footnote{Here the CNOT gate of control $c$ and target $t$ is $C_{c,t} = \frac12(I+Z)_c\otimes I_t + \frac12(I-Z)_c\otimes X_t$, we use $C=C_{1,2}$, and the Hadamard gate is $H = \frac{1}{\sqrt{2}}(X+Z)$.}
We also need to swap the two qubits in the middle by the swap gate $S = S_{(1,2)} = C_{1,2}C_{2,1}C_{1,2}$,
which arranges the first and second qubits of the two copies next to each other to carry out the contraction.
So, after acting with $B^\dagger\otimes B^\dagger$ and
performing measurements on each qubit in the computational basis, 
the $p_{0000}$ probability of the outcome $0000$ is proportional to the concurrence squared,
\begin{subequations}
\begin{equation}
\label{eq:p22B.c0}
    p_{0000} = \abs{\bra{0000}(B^\dagger\otimes B^\dagger) (I\otimes S \otimes I) (Y\otimes Y\otimes I\otimes I) (U_\psi\otimes U_\psi)\ket{0000}}^2
    = c^2(\psi)/2^2.
\end{equation}
We may simplify this circuit slightly.
Taking into account the identity
$\ket{\text{Bell}_\mu} = (\sigma_\mu\otimes I)\ket{\text{Bell}_0}$,
where $\sigma_\mu = (I,X,Y,Z)_\mu$ are the Pauli operators,
we have that the effect of the local Pauli-Y operator followed by the index contraction with $\bra{\text{Bell}_0}$
is the same as using simply $\bra{\text{Bell}_2} = \bra{\text{Bell}_0}(Y\otimes I)$.
So dropping the Pauli-Y gates and using the outcome $1111$ instead of $0000$, we end up with
\begin{equation}
\label{eq:p22B.c1}
    p_{1111} = \abs{\bra{1111}(B^\dagger\otimes B^\dagger) (I\otimes S \otimes I) (U_\psi\otimes U_\psi)\ket{0000}}^2
    = c^2(\psi)/2^2.
\end{equation}
Realizing that $B^\dagger = (H\otimes I)C$, we can get rid of the swap too, as
\begin{equation}
\label{eq:p22B.c}
    p_{1111} = \abs{\bra{1111} (H\otimes H\otimes I \otimes I) (C_{1,3}\otimes C_{2,4}) (U_\psi\otimes U_\psi)\ket{0000}}^2
    = c^2(\psi)/2^2.
\end{equation}
\end{subequations}
(The quantum circuits implementing this method are shown in the right part of Figure~\ref{fig:22.c}.)

We may realize that we can obtain the norm-squared also 
by using the state together with the complex conjugate of its copy,
and then contracting the respective indices by the Bell-measurement method.
That is, after preparing $\ket{\psi}\otimes\ket{\overline{\psi}} = (U_\psi\otimes\overline{U_\psi})\ket{0000}$
using the complex conjugate of $U_\psi$,
\begin{equation}
\label{eq:Upsi22Conj}
    \overline{U_\psi} := \sum_{i,j = 0}^1\ket{\overline{\chi}_{ij}}\bra{ij},
\end{equation}
we just have to do the index contraction by $\bra{\text{Bell}_0}$ as in the previous paragraph.
So, after acting with $B^\dagger\otimes B^\dagger$ and
performing measurements on each qubit in the computational basis,
the $p_{0000}$ probability of the outcome $0000$ is proportional to the square of the norm-square,
\begin{subequations}
\begin{equation}
\label{eq:p22B.n0}
    p_{0000} = \abs{\bra{0000}(B^\dagger\otimes B^\dagger) (I\otimes S \otimes I) (U_\psi\otimes \overline{U_\psi})\ket{0000}}^2
    = n^4(\psi)/2^2.
\end{equation}
Getting rid of the swap, we have
\begin{equation}
\label{eq:p22B.n}
    p_{0000} = \abs{\bra{0000}(H\otimes H\otimes I \otimes I) (C_{1,3}\otimes C_{2,4}) (U_\psi\otimes \overline{U_\psi})\ket{0000}}^2
    = n^4(\psi)/2^2.
\end{equation}
\end{subequations}
(The quantum circuits implementing this method are shown in the right part of Figure~\ref{fig:22.n}.)

Summarizing, we have two circuits for each invariant,
a smaller one, without Bell measurements,
and a larger one, with Bell measurements, using twice as many qubits.
The larger one has also the drawback that
any index contraction done by Bell measurements puts a factor $1/2$ to the probability (Eqs.~\eqref{eq:p22B.c0},~\eqref{eq:p22B.c},~\eqref{eq:p22B.n}),
increasing the relative error of the measurement.
Choosing the smaller one seems to be natural, however, sometimes the larger one may work better,
when $U_\psi^\T$ or $U_\psi^\dagger$ is harder to implement.
Note that if the state has real coefficients $\psi^{ij}$ only,
then one can choose a $U_\psi$ with real matrix elements,
and in this case $U_\psi$ can be used instead of $\overline{U_\psi}$, and $U_\psi^\dagger$ instead of $U_\psi^\T$.

\subsection{Three qubits}
\label{sec:theory.222}

Now let us turn to the case of three qubits.
For three-qubit state vectors $\ket{\psi} = \sum_{ijk = 0}^1\psi^{ijk}\ket{ijk}$
we have the following important invariants and covariants\footnote{From now on, summations for the repeated indices are understood.}~\cite{Borsten-2009,Szalay-2012b,Szalay-2013,Szalay-2025}
\begin{subequations}
\label{eq:222tens}
\begin{align}
    \label{eq:222tens.n}
    n^2(\psi) &:= \norm{\psi}^2 = \delta_{ii'}\delta_{jj'}\delta_{kk'}\overline{\psi^{ijk}}\psi^{i'j'k'},\\
    \label{eq:222tens.gam1}
    \ket{\gamma_1(\psi)} &:= \epsilon_{jj'}\epsilon_{kk'}\psi^{ijk}\psi^{i'j'k'}\ket{ii'},\\
    \label{eq:222tens.gam2}
    \ket{\gamma_2(\psi)} &:= \epsilon_{kk'}\epsilon_{ii'}\psi^{ijk}\psi^{i'j'k'}\ket{jj'},\\
    \label{eq:222tens.gam3}
    \ket{\gamma_3(\psi)} &:= \epsilon_{ii'}\epsilon_{jj'}\psi^{ijk}\psi^{i'j'k'}\ket{kk'},\\
    \label{eq:222tens.T}
    \ket{T(\psi)} &:= -\epsilon_{ll'}\epsilon_{mm'}\epsilon_{nn'}\psi^{imn}\psi^{lm'n'}\psi^{l'jk}\ket{ijk},\\
    \label{eq:222tens.q}
    q(\psi) &:= -2\Det(\psi) = 
    \epsilon_{ii'}\epsilon_{jj'}\epsilon_{kk'}\epsilon_{ll'}\epsilon_{mm'}\epsilon_{nn'}
    \psi^{ikl}\psi^{jk'l'}\psi^{i'mn}\psi^{j'm'n'}.
\end{align}
\end{subequations}
(See Appendix~\ref{sec:explicit.gen} for the explicit form for general states,
and Appendix~\ref{sec:explicit.canon} for the LU-canonical form~\cite{Acin-2000,Acin-2001}.)
Here $n^2(\psi)$ is $\mathrm{U}(8)$ invariant, 
and $\Det(\psi)$ is Cayley's hyperdeterminant~\cite{Cayley-1845,Gelfand-2008}, which is an $\mathrm{SL}(2)^{\times3}$ invariant.
The tensors $\ket{\gamma_1(\psi)}$, $\ket{\gamma_2(\psi)}$, $\ket{\gamma_3(\psi)}$, being symmetric,
transform as $(\mathbf{3},\mathbf{1},\mathbf{1})$, $(\mathbf{1},\mathbf{3},\mathbf{1})$, $(\mathbf{1},\mathbf{1},\mathbf{3})$,
and $\ket{T(\psi)}$ transforms as $(\mathbf{2},\mathbf{2},\mathbf{2})$ under $\mathrm{SL}(2)^{\times3}$.
On the other hand, $n^2(\psi)$ and $q(\psi)$ are invariant, and $\ket{T(\psi)}$ is covariant with respect to the permutation of the subsystems.
(This can be seen by the explicit form of these, see in Appendix~\ref{sec:explicit.gen}.
Accordingly, one may write two more equivalent definitions for $\ket{T(\psi)}$ and $q(\psi)$,
where the second or the third subsystem plays the role singled out in the contractions~\cite{Borsten-2009,Szalay-2025}.)
Subsystem permutation acts on the index $a$ of $\ket{\gamma_a(\psi)}$.

From the invariants and covariants~\eqref{eq:222tens} we again form $\mathrm{LU}$ invariants~\cite{Szalay-2012b,Szalay-2013}
\begin{subequations}
\label{eq:222invs}
\begin{align}
    \label{eq:222invs.n}
    n^4(\psi)      &:= \norm{\psi}^4,\\
    \label{eq:222invs.y}
    \begin{split}
    y^2(\psi)      &:= \frac23\bigl(g_1^2(\psi) + g_2^2(\psi) + g_3^2(\psi)\bigr) \\
                   &\phantom{:}= \frac13 \bigl(c_1^2(\psi) + c_2^2(\psi) + c_3^2(\psi)\bigr),
    \end{split}\\
    \label{eq:222invs.ca}
    c^2_a(\psi)    &:= g_b^2(\psi) + g_c^2(\psi) = 4\det(\rho_a),\\
    \label{eq:222invs.ga}
    g_a^2(\psi)    &:= \norm{\gamma_a(\psi)}^2,\\
    \label{eq:222invs.omega}
    \omega^2(\psi) &:= 4\norm{T(\psi)}^2,\\
    \label{eq:222invs.tau}
    \tau^2(\psi)   &:= 4\abs{q(\psi)}^2 = 16\abs{\Det(\psi)}^2.
\end{align}
\end{subequations}
(The explicit form of these would be too lengthy to write out for general states,
see Appendix~\ref{sec:explicit.canon} for the LU-canonical form~\cite{Acin-2000,Acin-2001}.)
Here $a$, $b$ and $c$ denote any elementary subsystem ($1$, $2$ or $3$),
which are always different in a given formula.
The constant prefactors are chosen so that
the maximal value of all of these invariants is the appropriate power of $\norm{\psi}=n(\psi)$,
so all of these take value between $0$ and $1$ in quantum probability~\cite{Szalay-2012b,Szalay-2013,Szalay-2025}.
(Note that here we define some of the invariants with different exponents as given in Ref.~\cite{Szalay-2012b,Szalay-2013},
as this turned out to be more natural~\cite{Szalay-2025}.
Also, $\omega^2$ is related to the Kempe invariant~\cite{Kempe-1999,Szalay-2025}, and it also appears in the twistor-geometric approach of three-qubit entanglement~\cite{Levay-2005,Szalay-2025}.)
These invariants or the powers of these in some cases, are measurable by our methods,
and are of direct importance in entanglement theory.

On the one hand, 
$c^2_a(\psi)=c_{a|bc}^2(\psi)$ is the \emph{concurrence squared}, the \emph{linear entropy of the reduced state} $\rho_a:=\Tr_{bc}(\ket{\psi}\bra{\psi})$, an \emph{entanglement measure}
quantifying the entanglement with respect to the $a|bc$ split of the system~\cite{Coffman-2000}.
Also, $\tau$ is the famous \emph{three-tangle}, an entanglement measure~\cite{Dur-2000}
quantifying the \emph{residual entanglement},
which is the tripartite entanglement that cannot be accounted for by the entanglement inside the two-qubit subsystems~\cite{Coffman-2000}.
A recent result is that $\omega$ is also an \emph{entanglement measure},
and the entanglement measures $c_a$, $\omega$ and $\tau$
together are \emph{ordered}, $0\leq\tau\leq\omega\leq c_a\leq 1$ for normalized vectors.
They measure $a|bc$-, W- and GHZ-entanglement, which are stronger and stronger in this order~\cite{Szalay-2025}.
Note that neither $\tau^2$, nor $\omega^2$ is an entanglement measure~\cite{Szalay-2025}.
Note also that, although $g_a$ is not an entanglement measure~\cite{Szalay-2012b,Szalay-2013,Szalay-2025},
it is still relevant as an indicator function used in the entanglement theory of \emph{mixed} three-qubit states~\cite{Szalay-2012b,Szalay-2013}.

On the other hand,
the invariants~\eqref{eq:222invs} originate in the FTS approach of three-qubit entanglement~\cite{Borsten-2009,Borsten-2013,Szalay-2012b,Szalay-2013,Szalay-2025},
which itself originates in the famous black hole/qubit correspondence~\cite{Borsten-2009b,Levay-2011b,Borsten-2012}
with direct connection to entanglement theory~\cite{Levay-2010b,Levay-2011a,Levay-2008,Vrana-2009,Sarosi-2014a,Sarosi-2014b,Sarosi-2014c,Holweck-2016}.
Here the tensors~\eqref{eq:222tens} are the natural objects,
and they identify the SLOCC classes of three-qubit entanglement~\cite{Dur-2000} (see in Table~\ref{tab:222classes})
in parallel to the FTS-rank, identified by the tensors~\eqref{eq:222tens}~\cite{Borsten-2009}.
Recall that the SLOCC classification is coarser than the LOCC classification~\cite{Dur-2000}, which we do not consider here.
In our case of three qubits, the former is discrete and SLOCC classes can be identified by the vanishing of the LSL-tensors~\eqref{eq:222tens}~\cite{Borsten-2009},
while the continuously many LOCC classes can be labeled by the values of an appropriate set of LU-invariants.
The vanishing of LSL-tensors is equivalent to the vanishing of their norms, which turn out to be LU-invariants~\eqref{eq:222invs}.
Note that, although we consider these LU-invariants, the SLOCC classification is still formulated by their vanishing, not their concrete values.

\begin{table}
\caption{SLOCC classes of three-qubit state vectors
identified by the vanishing of the invariants given in~\eqref{eq:222invs}.}
\label{tab:222classes}
\renewcommand{\arraystretch}{1.25}
\begin{tabular}{c||c|c|ccc|ccc|c|c}
    Class  & $n(\psi)$ & $y^2(\psi)$ & 
    $c^2_1(\psi)$ & $c^2_2(\psi)$ & $c^2_3(\psi)$ &
    $g_1^2(\psi)$ & $g_2^2(\psi)$ & $g_3^2(\psi)$ &
    $\omega^2(\psi)$ &
    $\tau^2(\psi)$  \\
    \hline
    \hline
    Null     & $=0$     & $=0$     & $=0$ & $=0$ & $=0$     & $=0$ & $=0$ & $=0$     & $=0$     & $=0$ \\
    \hline
    $1|2|3$  & $>0$     & $=0$     & $=0$ & $=0$ & $=0$     & $=0$ & $=0$ & $=0$     & $=0$     & $=0$ \\
    \hline
    $1|23$   & $>0$     & $>0$     & $=0$ & $>0$ & $>0$     & $>0$ & $=0$ & $=0$     & $=0$     & $=0$ \\
    $2|13$   & $>0$     & $>0$     & $>0$ & $=0$ & $>0$     & $=0$ & $>0$ & $=0$     & $=0$     & $=0$ \\
    $3|12$   & $>0$     & $>0$     & $>0$ & $>0$ & $=0$     & $=0$ & $=0$ & $>0$     & $=0$     & $=0$ \\
    \hline
    W        & $>0$     & $>0$     & $>0$ & $>0$ & $>0$     & $>0$ & $>0$ & $>0$     & $>0$     & $=0$ \\
    GHZ      & $>0$     & $>0$     & $>0$ & $>0$ & $>0$     & $>0$ & $>0$ & $>0$     & $>0$     & $>0$ \\
\end{tabular}
\end{table}%

For the quantum circuits implementing the measurement of the invariants~\eqref{eq:222invs},
the basic building blocks are the same as in the two-qubit case.
We need the unitary $U_\psi$ for the preparation of the state,
\begin{equation}
\label{eq:Upsi222}
    U_\psi := \sum_{i,j,k = 0}^1\ket{\chi_{ijk}}\bra{ijk},
\end{equation}
(where $\ket{\chi_{ijk}}$ are orthonormal, and $\ket{\chi_{000}} := \ket{\psi}$),
together with its adjoint $U_\psi^\dagger$, transpose $U_\psi^\T$ and conjugate $\overline{U_\psi}$,
defined analogously to those of the two-qubit case in~\eqref{eq:Upsi22},~\eqref{eq:Upsi22Adj},~\eqref{eq:Upsi22Transp} and~\eqref{eq:Upsi22Conj}.

\begin{figure}
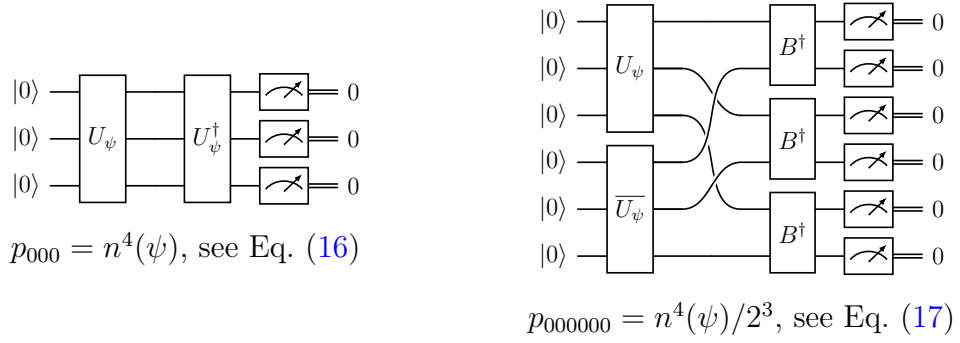

\begin{minipage}{0.48\textwidth}
\centering
    \myquantikz{
    \lstick{$\ket{0}$} & \gate[3,\DAH]{U_\psi} & & \gate[3,\DAH]{U_\psi^\dagger}
     & \meter{} & \setwiretype{c} \rstick{$0$}  \\
    \lstick{$\ket{0}$} &    &    &
     & \meter{} & \setwiretype{c} \rstick{$0$}  \\
    \lstick{$\ket{0}$} &    &    &
     & \meter{} & \setwiretype{c} \rstick{$0$}
    }\\
    $p_{000} = n^4(\psi)$, see Eq.~\eqref{eq:p222.n} 
\end{minipage}
\begin{minipage}{0.48\textwidth}
\centering
    \myquantikz{
    \lstick{$\ket{0}$} & \gate[3,\DAH]{U_\psi} &    & \gate[2,\DAH]{B^\dagger}
     & \meter{} & \setwiretype{c} \rstick{$0$}  \\
    \lstick{$\ket{0}$} &    & \permute{2,4,1,3} &
     & \meter{} & \setwiretype{c} \rstick{$0$}  \\
    \lstick{$\ket{0}$} &    &    & \gate[2,\DAH]{B^\dagger}
     & \meter{} & \setwiretype{c} \rstick{$0$}  \\
    \lstick{$\ket{0}$} & \gate[3,\DAH]{\overline{U_\psi}} &    & 
     & \meter{} & \setwiretype{c} \rstick{$0$}  \\
    \lstick{$\ket{0}$} &    &    & \gate[2,\DAH]{B^\dagger}
     & \meter{} & \setwiretype{c} \rstick{$0$}  \\
    \lstick{$\ket{0}$} &    &    &
     & \meter{} & \setwiretype{c} \rstick{$0$}
    } \\  
    $p_{000000} = n^4(\psi)/2^3$, see Eq.~\eqref{eq:p222B.n}
\end{minipage}
\caption{Quantum circuits implementing the three-qubit norm-squared \eqref{eq:222invs.n}.}
\label{fig:222.n}
\end{figure}

Let us start with the norm-squared $n^2(\psi)$ in \eqref{eq:222invs.n}.
We have two possibilities, similarly to the two-qubit case.
The smaller circuit (shown in the left part of Figure~\ref{fig:222.n})
uses three qubits and $U_\psi$ and $U_\psi^\dagger$ once, and gives
\begin{equation}
\label{eq:p222.n}
    p_{000} = \abs{\bra{000}U_\psi^\dagger U_\psi\ket{000}}^2 = n^4(\psi).
\end{equation}
The larger circuit (shown in the right part of Figure~\ref{fig:222.n})
uses six qubits and one instance of $U_\psi$ and $\overline{U_\psi}$
and also three instances of the Bell gate $B^\dagger$ with the appropriate permutation, or with CNOTs,
and gives
\begin{equation}
\label{eq:p222B.n}
\begin{split}
    p_{000000} &= \abs{\bra{000000}B^{\dagger\otimes3}
    (S_{(2,3,5,4)}\otimes I_{1,6})
    (U_\psi\otimes \overline{U_\psi})\ket{000000}}^2 \\
    &= \abs{\bra{000000}(H^{\otimes3}\otimes I^{\otimes3})
    (C_{1,4}\otimes C_{2,5}\otimes C_{3,6})
    (U_\psi\otimes \overline{U_\psi})\ket{000000}}^2 \\
    &= n^4(\psi)/2^3.
\end{split}
\end{equation}
(In this section we draw only the permutation variant of the larger circuits,
since it is much more expressive,
and the CNOT variant would take up too much space, especially later.
Translating the former to the latter is straightforward.)

\begin{figure}
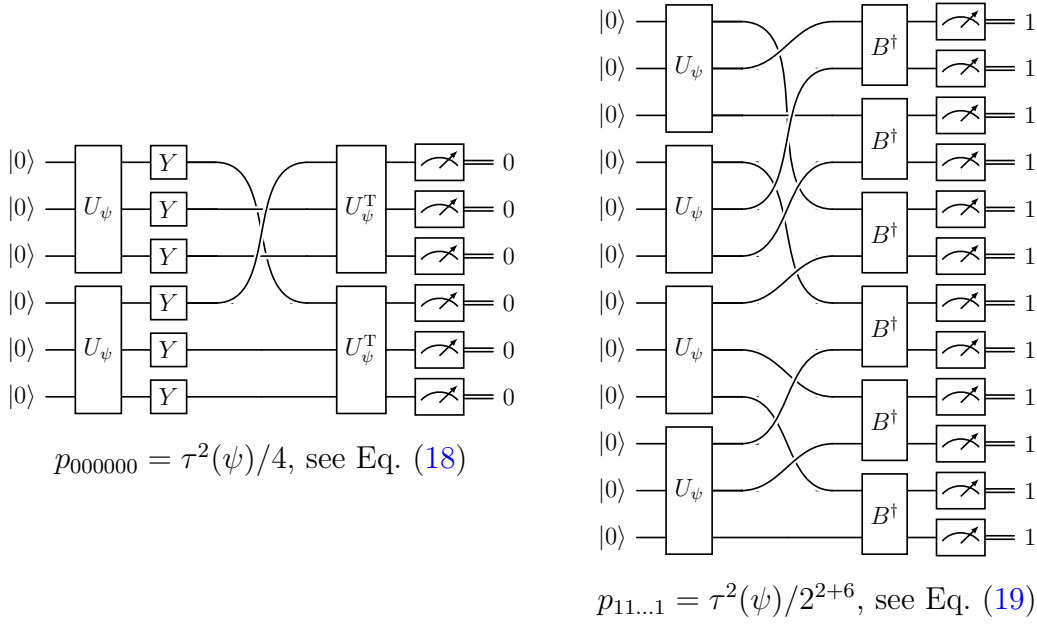

\begin{minipage}{0.48\textwidth}
\centering
    \myquantikz{
    \lstick{$\ket{0}$} & \gate[3,\DAH]{U_\psi} & \gate{Y} &  \permute{4,2,3,1}\hphantom{iiii}  & \gate[3,\DAH]{U_\psi^\T}
     & \meter{} & \setwiretype{c} \rstick{$0$}  \\
    \lstick{$\ket{0}$} &    & \gate{Y} &    &
     & \meter{} & \setwiretype{c} \rstick{$0$}  \\
    \lstick{$\ket{0}$} &    & \gate{Y} &    &
     & \meter{} & \setwiretype{c} \rstick{$0$}  \\
    \lstick{$\ket{0}$} & \gate[3,\DAH]{U_\psi} & \gate{Y} &    & \gate[3,\DAH]{U_\psi^\T}
     & \meter{} & \setwiretype{c} \rstick{$0$}  \\
    \lstick{$\ket{0}$} &    & \gate{Y} &    &
     & \meter{} & \setwiretype{c} \rstick{$0$}  \\
    \lstick{$\ket{0}$} &    & \gate{Y} &    &
     & \meter{} & \setwiretype{c} \rstick{$0$}
    }\\
    $p_{000000} = \tau^2(\psi)/4$, see Eq.~\eqref{eq:p222.tau} 
\end{minipage}
\begin{minipage}{0.48\textwidth}
\centering
    \myquantikz{
    \lstick{$\ket{0}$} & \gate[3,\DAH]{U_\psi} & \permute{5,1,3,7,2,4,6,9,11,8,10}\hphantom{iiii} & \gate[2,\DAH]{B^\dagger}
     & \meter{} & \setwiretype{c} \rstick{$1$}  \\
    \lstick{$\ket{0}$} &    &    &
     & \meter{} & \setwiretype{c} \rstick{$1$}  \\
    \lstick{$\ket{0}$} &    &    & \gate[2,\DAH]{B^\dagger}
     & \meter{} & \setwiretype{c} \rstick{$1$}  \\
    \lstick{$\ket{0}$} & \gate[3,\DAH]{U_\psi} &    & 
     & \meter{} & \setwiretype{c} \rstick{$1$}  \\
    \lstick{$\ket{0}$} &    &    & \gate[2,\DAH]{B^\dagger}
     & \meter{} & \setwiretype{c} \rstick{$1$}  \\
    \lstick{$\ket{0}$} &    &    &
     & \meter{} & \setwiretype{c} \rstick{$1$}  \\
    \lstick{$\ket{0}$} & \gate[3,\DAH]{U_\psi} &    & \gate[2,\DAH]{B^\dagger}
     & \meter{} & \setwiretype{c} \rstick{$1$}  \\
    \lstick{$\ket{0}$} &    &    &
     & \meter{} & \setwiretype{c} \rstick{$1$}  \\
    \lstick{$\ket{0}$} &    &    & \gate[2,\DAH]{B^\dagger}
     & \meter{} & \setwiretype{c} \rstick{$1$}  \\
    \lstick{$\ket{0}$} & \gate[3,\DAH]{U_\psi} &    & 
     & \meter{} & \setwiretype{c} \rstick{$1$}  \\
    \lstick{$\ket{0}$} &    &    & \gate[2,\DAH]{B^\dagger}
     & \meter{} & \setwiretype{c} \rstick{$1$}  \\
    \lstick{$\ket{0}$} &    &    &
     & \meter{} & \setwiretype{c} \rstick{$1$}
    }\\
    $p_{11\dots1} = \tau^2(\psi)/2^{2+6}$, see Eq.~\eqref{eq:p222B.tau}
\end{minipage}
\caption{Quantum circuits implementing the three-tangle \eqref{eq:222invs.tau}.}
\label{fig:222.tau}
\end{figure}

The three-tangle $\tau(\psi)$ in \eqref{eq:222invs.tau} can also be implemented in two ways.
The index contraction scheme can be read off from~\eqref{eq:222tens.q},
in the scalar $q(\psi)$ there are contractions with $\epsilon$.
The smaller circuit (shown in the left part of Figure~\ref{fig:222.tau})
uses six qubits and two instances of $U_\psi$ and $U_\psi^\T$, a swap, and gives
\begin{equation}
\label{eq:p222.tau}
\begin{split}
    p_{000000} &= \abs{ \bra{000000} (U_\psi^{\T\otimes2}) (S_{(1,4)}\otimes I_{2,3,5,6}) (Y^{\otimes6}) (U_\psi^{\otimes2}) \ket{000000} }^2\\
    &= \abs{q(\psi)}^2 = \tau^2(\psi)/2^2.
\end{split}
\end{equation}
Note that $(S_{(2,5)}\otimes I_{1,3,4,6})$ or $(S_{(3,6)}\otimes I_{1,2,4,5})$ could also be used,
because of the permutation invariance of $q(\psi)$.
The larger circuit (shown in the right part of Figure~\ref{fig:222.tau})
uses twelve qubits and four instances of $U_\psi$, and also six instances of the Bell gate $B^\dagger$
with the appropriate permutation, or with CNOTs, and gives
\begin{equation}
\label{eq:p222B.tau}
\begin{split}
    p_{11\dots1} &= \abs{ \bra{11\dots1} (B^{\dagger\otimes6})(S_\sigma)
    (U_\psi^{\otimes4}) \ket{00\dots0} }^2\\
    &= \abs{ \bra{11\dots1} (H^{\otimes4}\otimes I^{\otimes3}\otimes H^{\otimes2}\otimes I^{\otimes3})(C_s)(U_\psi^{\otimes4}) \ket{00\dots0} }^2 \\
    &= \abs{q(\psi)}^2/2^6 = \tau^2(\psi)/2^{2+6},
\end{split}
\end{equation}
where $S_\sigma$ is the unitary permutation operator implementing the permutation
$\sigma=(1,5,2)(3)(4,7,6)(8,9,11,10)(12)$,
and the CNOTs are $C_s=C_{1,7}\otimes C_{2,5}\otimes C_{3,6}\otimes C_{4,10}\otimes C_{8,11}\otimes C_{9,12}$.

\begin{figure}
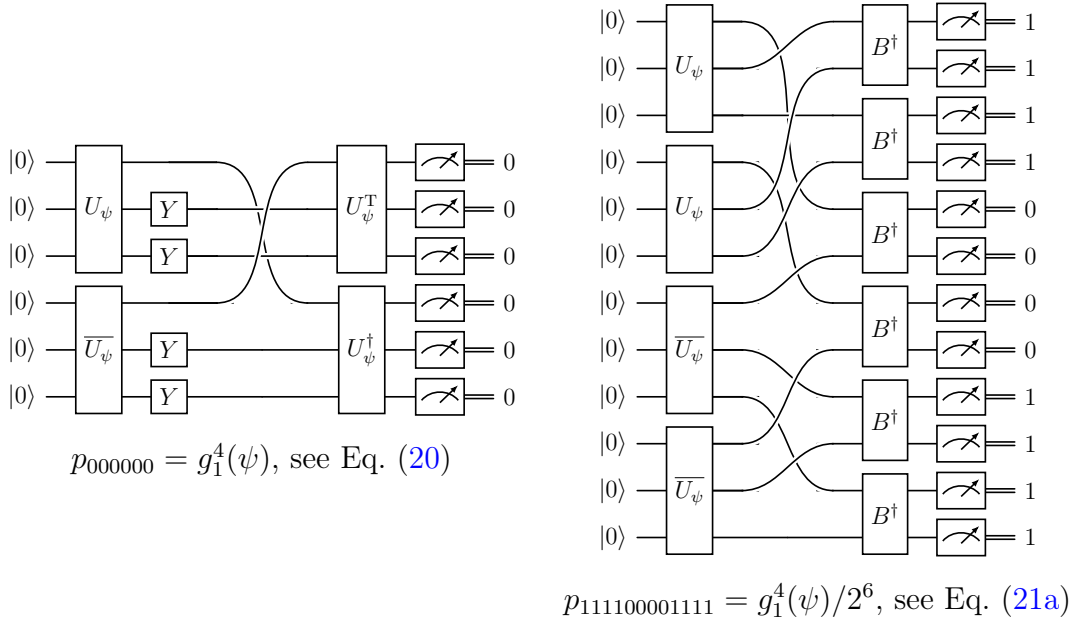

\begin{minipage}{0.48\textwidth}
\centering
    \myquantikz{
    \lstick{$\ket{0}$} & \gate[3,\DAH]{U_\psi} &    &  \permute{4,2,3,1}\hphantom{iiii}  & \gate[3,\DAH]{U_\psi^\T}
     & \meter{} & \setwiretype{c} \rstick{$0$}  \\
    \lstick{$\ket{0}$} &    & \gate{Y} &    &
     & \meter{} & \setwiretype{c} \rstick{$0$}  \\
    \lstick{$\ket{0}$} &    & \gate{Y} &    &
     & \meter{} & \setwiretype{c} \rstick{$0$}  \\
    \lstick{$\ket{0}$} & \gate[3,\DAH]{\overline{U_\psi}} &    &    & \gate[3,\DAH]{U_\psi^\dagger}
     & \meter{} & \setwiretype{c} \rstick{$0$}  \\
    \lstick{$\ket{0}$} &    & \gate{Y} &    &
     & \meter{} & \setwiretype{c} \rstick{$0$}  \\
    \lstick{$\ket{0}$} &    & \gate{Y} &    &
     & \meter{} & \setwiretype{c} \rstick{$0$}
    }\\
    $p_{000000} = g_1^4(\psi)$, see Eq.~\eqref{eq:p222.g1} 
\end{minipage}
\begin{minipage}{0.48\textwidth}
\centering
    \myquantikz{
    \lstick{$\ket{0}$} & \gate[3,\DAH]{U_\psi} & \permute{5,1,3,7,2,4,6,9,11,8,10}\hphantom{iiii} & \gate[2,\DAH]{B^\dagger}
     & \meter{} & \setwiretype{c} \rstick{$1$}  \\
    \lstick{$\ket{0}$} &    &    &
     & \meter{} & \setwiretype{c} \rstick{$1$}  \\
    \lstick{$\ket{0}$} &    &    & \gate[2,\DAH]{B^\dagger}
     & \meter{} & \setwiretype{c} \rstick{$1$}  \\
    \lstick{$\ket{0}$} & \gate[3,\DAH]{U_\psi} &    & 
     & \meter{} & \setwiretype{c} \rstick{$1$}  \\
    \lstick{$\ket{0}$} &    &    & \gate[2,\DAH]{B^\dagger}
     & \meter{} & \setwiretype{c} \rstick{$0$}  \\
    \lstick{$\ket{0}$} &    &    &
     & \meter{} & \setwiretype{c} \rstick{$0$}  \\
    \lstick{$\ket{0}$} & \gate[3,\DAH]{\overline{U_\psi}} &    & \gate[2,\DAH]{B^\dagger}
     & \meter{} & \setwiretype{c} \rstick{$0$}  \\
    \lstick{$\ket{0}$} &    &    &
     & \meter{} & \setwiretype{c} \rstick{$0$}  \\
    \lstick{$\ket{0}$} &    &    & \gate[2,\DAH]{B^\dagger}
     & \meter{} & \setwiretype{c} \rstick{$1$}  \\
    \lstick{$\ket{0}$} & \gate[3,\DAH]{\overline{U_\psi}} &    & 
     & \meter{} & \setwiretype{c} \rstick{$1$}  \\
    \lstick{$\ket{0}$} &    &    & \gate[2,\DAH]{B^\dagger}
     & \meter{} & \setwiretype{c} \rstick{$1$}  \\
    \lstick{$\ket{0}$} &    &    &
     & \meter{} & \setwiretype{c} \rstick{$1$}
    }\\
    $p_{111100001111} = g_1^4(\psi)/2^6$, see Eq.~\eqref{eq:p222B.g1}
\end{minipage}
\caption{Quantum circuits implementing the three-qubit $g_1(\psi)$ \eqref{eq:222invs.ga}.}
\label{fig:222.g1}
\end{figure}

The invariant $g_a^2(\psi)$ in \eqref{eq:222invs.ga} can also be implemented in two ways.
The index contraction scheme can be read off from~\eqref{eq:222tens.gam1} and~\eqref{eq:222invs.ga},
in the tensor $\ket{\gamma_a(\psi)}$ there are contractions with $\epsilon$,
then in the norm 
$\norm{\gamma_a(\psi)}^2=\delta_{ii'}\delta_{ll'}\overline{\gamma_a(\psi)^{il}}\gamma_a(\psi)^{i'l'}$ 
there are contractions with $\delta$.
The smaller circuit (shown in the left part of Figure~\ref{fig:222.g1})
uses six qubits and $U_\psi$, $\overline{U_\psi}$, $U_\psi^\T$ and $U_\psi^\dagger$, a swap, and gives
\begin{equation}
\label{eq:p222.g1}
\begin{split}
    &p_{000000} \\
    &\quad = \abs{\bra{000000}(U_\psi^\T\otimes U_\psi^\dagger) (S_{(1,4)}\otimes I_{2,3,5,6}) (I\otimes Y\otimes Y)^{\otimes2} (U_\psi\otimes \overline{U_\psi})\ket{000000}}^2\\
    &\quad = g_1^4(\psi).
\end{split}
\end{equation}
Note that $(S_{(2,5)}\otimes I_{1,3,4,6})(Y\otimes I\otimes Y)^{\otimes2}$
or $(S_{(3,6)}\otimes I_{1,2,4,5})(Y\otimes Y\otimes I)^{\otimes2}$ are to be used for $g_2^2(\psi)$ and $g_3^2(\psi)$, respectively,
because of the permutation covariance of $g_a^2(\psi)$ in the index $a$.
The larger circuit (shown in the right part of Figure~\ref{fig:222.g1})
uses twelve qubits and two instances of $U_\psi$ and $\overline{U_\psi}$, and also six instances of the Bell gate $B^\dagger$
with the appropriate permutation, or with CNOTs, and gives
\begin{subequations}
\begin{align}
\label{eq:p222B.g1}
\begin{split}
    &p_{111100001111} \\
    &\quad = \abs{ \bra{111100001111} (B^{\dagger\otimes6})(S_\sigma)
    (U_\psi^{\otimes2}\otimes\overline{U_\psi}^{\otimes2}) \ket{00\dots0} }^2\\
    &\quad = g_1(\psi)^4/2^6,
\end{split}\\
\begin{split}
    &p_{011011011011} \\
    &\quad = \abs{ \bra{011011011011} (H^{\otimes4}\otimes I^{\otimes3}\otimes H^{\otimes2}\otimes I^{\otimes3})\\
    &\quad \qquad (C_s)(U_\psi^{\otimes2}\otimes\overline{U_\psi}^{\otimes2}) \ket{00\dots0} }^2\\
    &\quad = g_1(\psi)^4/2^6.
\end{split}
\end{align}
\end{subequations}
Note that in the two variants of the circuit different outcomes give the result,
since there are contractions by both $\epsilon$ and $\delta$ here.

\begin{figure}
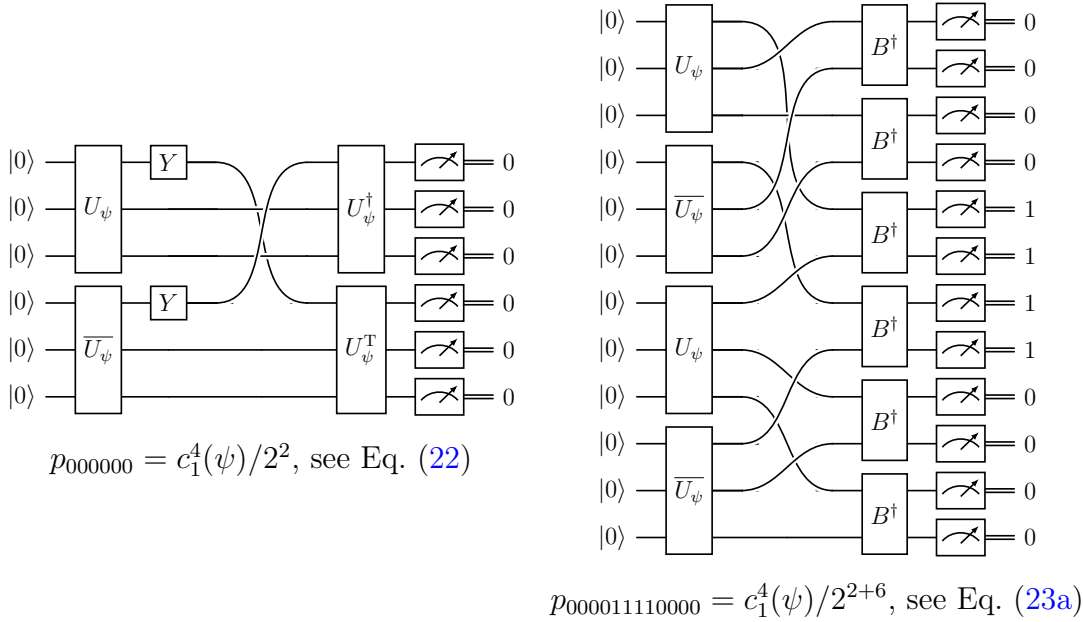

\begin{minipage}{0.48\textwidth}
\centering
    \myquantikz{
    \lstick{$\ket{0}$} & \gate[3,\DAH]{U_\psi} & \gate{Y} &  \permute{4,2,3,1}\hphantom{iiii}  & \gate[3,\DAH]{U_\psi^\dagger}
     & \meter{} & \setwiretype{c} \rstick{$0$}  \\
    \lstick{$\ket{0}$} &    &    &    &
     & \meter{} & \setwiretype{c} \rstick{$0$}  \\
    \lstick{$\ket{0}$} &    &    &    &
     & \meter{} & \setwiretype{c} \rstick{$0$}  \\
    \lstick{$\ket{0}$} & \gate[3,\DAH]{\overline{U_\psi}} & \gate{Y} &    & \gate[3,\DAH]{U_\psi^\T}
     & \meter{} & \setwiretype{c} \rstick{$0$}  \\
    \lstick{$\ket{0}$} &    &    &    &
     & \meter{} & \setwiretype{c} \rstick{$0$}  \\
    \lstick{$\ket{0}$} &    &    &    &
     & \meter{} & \setwiretype{c} \rstick{$0$}
    }\\
    $p_{000000} = c_1^4(\psi)/2^2$, see Eq.~\eqref{eq:p222.c1} 
\end{minipage}
\begin{minipage}{0.48\textwidth}
\centering
    \myquantikz{
    \lstick{$\ket{0}$} & \gate[3,\DAH]{U_\psi} & \permute{5,1,3,7,2,4,6,9,11,8,10}\hphantom{iiii} & \gate[2,\DAH]{B^\dagger}
     & \meter{} & \setwiretype{c} \rstick{$0$}  \\
    \lstick{$\ket{0}$} &    &    &
     & \meter{} & \setwiretype{c} \rstick{$0$}  \\
    \lstick{$\ket{0}$} &    &    & \gate[2,\DAH]{B^\dagger}
     & \meter{} & \setwiretype{c} \rstick{$0$}  \\
    \lstick{$\ket{0}$} & \gate[3,\DAH]{\overline{U_\psi}} &    & 
     & \meter{} & \setwiretype{c} \rstick{$0$}  \\
    \lstick{$\ket{0}$} &    &    & \gate[2,\DAH]{B^\dagger}
     & \meter{} & \setwiretype{c} \rstick{$1$}  \\
    \lstick{$\ket{0}$} &    &    &
     & \meter{} & \setwiretype{c} \rstick{$1$}  \\
    \lstick{$\ket{0}$} & \gate[3,\DAH]{U_\psi} &    & \gate[2,\DAH]{B^\dagger}
     & \meter{} & \setwiretype{c} \rstick{$1$}  \\
    \lstick{$\ket{0}$} &    &    &
     & \meter{} & \setwiretype{c} \rstick{$1$}  \\
    \lstick{$\ket{0}$} &    &    & \gate[2,\DAH]{B^\dagger}
     & \meter{} & \setwiretype{c} \rstick{$0$}  \\
    \lstick{$\ket{0}$} & \gate[3,\DAH]{\overline{U_\psi}} &    & 
     & \meter{} & \setwiretype{c} \rstick{$0$}  \\
    \lstick{$\ket{0}$} &    &    & \gate[2,\DAH]{B^\dagger}
     & \meter{} & \setwiretype{c} \rstick{$0$}  \\
    \lstick{$\ket{0}$} &    &    &
     & \meter{} & \setwiretype{c} \rstick{$0$}
    }\\
    $p_{000011110000} = c_1^4(\psi)/2^{2+6}$, see Eq.~\eqref{eq:p222B.c1}
\end{minipage}
\caption{Quantum circuits implementing the three-qubit local concurrence $c_1(\psi)$ \eqref{eq:222invs.ca}.}
\label{fig:222.c1}
\end{figure}

The invariant $c_a^2(\psi)$ in \eqref{eq:222invs.ca} can also be implemented in two ways directly,
without calculating the sum of $g_b^2(\psi)$ and $g_c^2(\psi)$ coming from two different circuits shown before.
The index contraction scheme can be read off from the second form in~\eqref{eq:222invs.ca},
in the partial trace $\rho_a^{ii'}=\Tr_{bc}(\ket{\psi}\bra{\psi})^{ii'}
=\delta_{jj'}\delta_{kk'}\overline{\psi^{ijk}}\psi^{i'j'k'}$ there are contractions with $\delta$,
then in the determinant $2\det(\rho_a)=\epsilon_{ii'}\epsilon_{ll'}\rho_a^{il}\rho_a^{i'l'}$ there are contractions with $\epsilon$.
The smaller circuit (shown in the left part of Figure~\ref{fig:222.c1})
uses six qubits and $U_\psi$, $\overline{U_\psi}$, $U_\psi^\T$ and $U_\psi^\dagger$, a swap, and gives
\begin{equation}
\label{eq:p222.c1}
\begin{split}
    &p_{000000} \\
    &\quad = \abs{\bra{000000}(U_\psi^\dagger\otimes U_\psi^\T) (S_{(1,4)}\otimes I_{2,3,5,6}) (Y\otimes I\otimes I)^{\otimes2} (U_\psi\otimes \overline{U_\psi})\ket{000000}}^2\\
    &\quad = c_1^4(\psi)/2^2.
\end{split}
\end{equation}
Note that $(S_{(2,5)}\otimes I_{1,3,4,6})(I\otimes Y\otimes I)^{\otimes2}$
or $(S_{(3,6)}\otimes I_{1,2,4,5})(I\otimes I\otimes Y)^{\otimes2}$ are to be used for $c_2^2(\psi)$ and $c_3^2(\psi)$, respectively,
because of the permutation covariance of $c_a^2(\psi)$ in the index $a$.
The larger circuit (shown in the right part of Figure~\ref{fig:222.c1})
uses twelve qubits and two instances of $U_\psi$ and $\overline{U_\psi}$, and also six instances of the Bell gate $B^\dagger$
with the appropriate permutation, and gives
\begin{subequations}
\begin{align}
\label{eq:p222B.c1}
\begin{split}
    &p_{000011110000} \\
    &\quad = \abs{ \bra{111100001111} (B^{\dagger\otimes6})(S_\sigma)
    (U_\psi\otimes\overline{U_\psi}\otimes U_\psi\otimes\overline{U_\psi}) \ket{00\dots0} }^2\\
    &\quad = c_1(\psi)^4/2^{2+6},
\end{split}\\
\begin{split}
    &p_{100100100100} \\
    &\quad = \abs{ \bra{100100100100} (H^{\otimes4}\otimes I^{\otimes3}\otimes H^{\otimes2}\otimes I^{\otimes3})\\
    &\quad \qquad (C_s)(U_\psi\otimes\overline{U_\psi}\otimes U_\psi\otimes\overline{U_\psi}) \ket{00\dots0} }^2\\
    &\quad = c_1(\psi)^4/2^{2+6}.
\end{split}
\end{align}
\end{subequations}

\begin{figure}
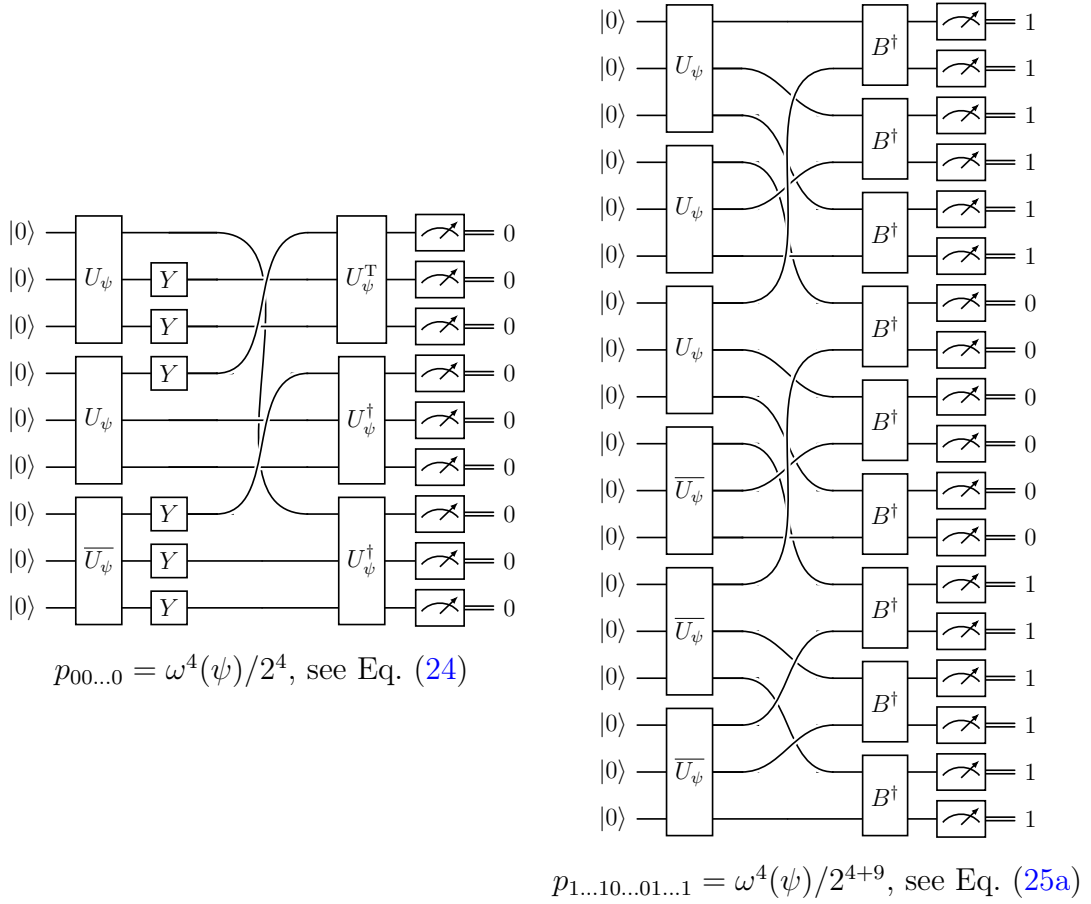

\begin{minipage}{0.48\textwidth}
\centering
    \myquantikz{
    \lstick{$\ket{0}$} & \gate[3,\DAH]{U_\psi} &    &  \permute{7,2,3,1,5,6,4}\hphantom{iiii}  & \gate[3,\DAH]{U_\psi^\T}
     & \meter{} & \setwiretype{c} \rstick{$0$}  \\
    \lstick{$\ket{0}$} &    & \gate{Y} &    &
     & \meter{} & \setwiretype{c} \rstick{$0$}  \\
    \lstick{$\ket{0}$} &    & \gate{Y} &    &
     & \meter{} & \setwiretype{c} \rstick{$0$}  \\
    \lstick{$\ket{0}$} & \gate[3,\DAH]{U_\psi} & \gate{Y} &    & \gate[3,\DAH]{U_\psi^\dagger}
     & \meter{} & \setwiretype{c} \rstick{$0$}  \\
    \lstick{$\ket{0}$} &    &    &    &
     & \meter{} & \setwiretype{c} \rstick{$0$}  \\
    \lstick{$\ket{0}$} &    &    &    &
     & \meter{} & \setwiretype{c} \rstick{$0$}  \\
    \lstick{$\ket{0}$} & \gate[3,\DAH]{\overline{U_\psi}} & \gate{Y} &    & \gate[3,\DAH]{U_\psi^\dagger}
     & \meter{} & \setwiretype{c} \rstick{$0$}  \\
    \lstick{$\ket{0}$} &    & \gate{Y} &    &
     & \meter{} & \setwiretype{c} \rstick{$0$}  \\
    \lstick{$\ket{0}$} &    & \gate{Y} &    &
     & \meter{} & \setwiretype{c} \rstick{$0$}
    }\\
    $p_{00\dots0} = \omega^4(\psi)/2^4$, see Eq.~\eqref{eq:p222.omega} 
\end{minipage}
\begin{minipage}{0.48\textwidth}
\centering
    \myquantikz{
    \lstick{$\ket{0}$} & \gate[3,\DAH]{U_\psi} &    & \gate[2,\DAH]{B^\dagger}
     & \meter{} & \setwiretype{c} \rstick{$1$}  \\
    \lstick{$\ket{0}$} &    & \permute{2,4,6,3,5,1,8,10,12,9,11,7,14,16,13,15}\hphantom{iiii} &
     & \meter{} & \setwiretype{c} \rstick{$1$}  \\
    \lstick{$\ket{0}$} &    &    & \gate[2,\DAH]{B^\dagger}
     & \meter{} & \setwiretype{c} \rstick{$1$}  \\
    \lstick{$\ket{0}$} & \gate[3,\DAH]{U_\psi} &    & 
     & \meter{} & \setwiretype{c} \rstick{$1$}  \\
    \lstick{$\ket{0}$} &    &    & \gate[2,\DAH]{B^\dagger}
     & \meter{} & \setwiretype{c} \rstick{$1$}  \\
    \lstick{$\ket{0}$} &    &    &
     & \meter{} & \setwiretype{c} \rstick{$1$}  \\
    \lstick{$\ket{0}$} & \gate[3,\DAH]{U_\psi} &    & \gate[2,\DAH]{B^\dagger}
     & \meter{} & \setwiretype{c} \rstick{$0$}  \\
    \lstick{$\ket{0}$} &    &    &
     & \meter{} & \setwiretype{c} \rstick{$0$}  \\
    \lstick{$\ket{0}$} &    &    & \gate[2,\DAH]{B^\dagger}
     & \meter{} & \setwiretype{c} \rstick{$0$}  \\
    \lstick{$\ket{0}$} & \gate[3,\DAH]{\overline{U_\psi}} &    & 
     & \meter{} & \setwiretype{c} \rstick{$0$}  \\
    \lstick{$\ket{0}$} &    &    & \gate[2,\DAH]{B^\dagger}
     & \meter{} & \setwiretype{c} \rstick{$0$}  \\
    \lstick{$\ket{0}$} &    &    &
     & \meter{} & \setwiretype{c} \rstick{$0$}  \\
    \lstick{$\ket{0}$} & \gate[3,\DAH]{\overline{U_\psi}} &    & \gate[2,\DAH]{B^\dagger}
     & \meter{} & \setwiretype{c} \rstick{$1$}  \\
    \lstick{$\ket{0}$} &    &    &
     & \meter{} & \setwiretype{c} \rstick{$1$}  \\
    \lstick{$\ket{0}$} &    &    & \gate[2,\DAH]{B^\dagger}
     & \meter{} & \setwiretype{c} \rstick{$1$}  \\
    \lstick{$\ket{0}$} & \gate[3,\DAH]{\overline{U_\psi}} &    & 
     & \meter{} & \setwiretype{c} \rstick{$1$}  \\
    \lstick{$\ket{0}$} &    &    & \gate[2,\DAH]{B^\dagger}
     & \meter{} & \setwiretype{c} \rstick{$1$}  \\
    \lstick{$\ket{0}$} &    &    &
     & \meter{} & \setwiretype{c} \rstick{$1$}    
    }\\
    $p_{1\dots10\dots01\dots1} = \omega^4(\psi)/2^{4+9}$, see Eq.~\eqref{eq:p222B.omega}
\end{minipage}
\caption{Quantum circuits implementing the three-qubit $\omega(\psi)$ \eqref{eq:222invs.omega}.}
\label{fig:222.omega}
\end{figure}

The invariant $\omega^2(\psi)$ in \eqref{eq:222invs.omega} can also be implemented in two ways.
The index contraction scheme can be read off from~\eqref{eq:222tens.T} and~\eqref{eq:222invs.omega},
in the tensor $\ket{T(\psi)}$ there are contractions with $\epsilon$,
then in the norm $\norm{T(\psi)}^2=\delta_{ii'}\delta_{jj'}\delta_{kk'}\overline{T(\psi)^{ijk}}T(\psi)^{i'j'k'}$ there are contractions with $\delta$.
The smaller circuit (shown in the left part of Figure~\ref{fig:222.omega})
uses nine qubits and two instances of $U_\psi$ and $\overline{U_\psi}$, one instance of $U_\psi^\T$ and $U_\psi^\dagger$ with the appropriate swaps (a particular decomposition of the permutation $S_{(1,4,7)}$), and gives
\begin{equation}
\label{eq:p222.omega}
\begin{split}
    &p_{000000} \\
    &\quad = \abs{\bra{000000}(U_\psi^\T\otimes U_\psi^\dagger\otimes U_\psi^\dagger) 
     ((S_{(1,4)}\otimes I_7)(S_{(4,7)}\otimes I_1)(S_{(7,1)}\otimes I_4))\otimes I_{2,3,5,6,8,9})\\
    &\quad \qquad (I\otimes Y\otimes Y\otimes Y\otimes I\otimes I\otimes Y\otimes Y\otimes Y) 
    (U_\psi\otimes U_\psi\otimes \overline{U_\psi})\ket{000000}}^2\\
    &\quad = \omega^4(\psi)/2^4.
\end{split}
\end{equation}
Note that 
$((S_{(2,5)}\otimes I_8)(S_{(5,8)}\otimes I_2)(S_{(8,2)}\otimes I_5))\otimes I_{1,3,4,6,7,9})
 (Y\otimes I\otimes Y\otimes I\otimes Y\otimes I\otimes Y\otimes Y\otimes Y)$
or
$((S_{(3,6)}\otimes I_9)(S_{(6,9)}\otimes I_3)(S_{(9,3)}\otimes I_6))\otimes I_{1,2,4,5,7,8})
 (Y\otimes Y\otimes I\otimes I\otimes I\otimes Y\otimes Y\otimes Y\otimes Y)$
can also be used,
because of the permutation invariance of $\omega^2(\psi)$.
The larger circuit (shown in the right part of Figure~\ref{fig:222.omega})
uses eighteen qubits and three instances of $U_\psi$ and $\overline{U_\psi}$, and also nine instances of the Bell gate $B^\dagger$
with the appropriate permutation, or CNOTS, and gives
\begin{subequations}
\begin{align}
\label{eq:p222B.omega}
\begin{split}
    &p_{111111000000111111} \\
    &\quad = \abs{ \bra{111111000000111111} (B^{\dagger\otimes9})(S_{\sigma'})
    (U_\psi^{\otimes3}\otimes \overline{U_\psi}^{\otimes3}) \ket{00\dots0} }^2\\
    &\quad = \omega(\psi)^4/2^{4+9},
\end{split}\\
\begin{split}
    &p_{111011100100011111} \\
    &\quad = \abs{ \bra{111011100100011111} 
    ( H^{\otimes4}\otimes I^{\otimes3}\otimes H^{\otimes3}\otimes I^{\otimes3}\otimes H^{\otimes2}\otimes I^{\otimes3})\\
    &\quad \qquad (C_{s'})(U_\psi^{\otimes3}\otimes \overline{U_\psi}^{\otimes3}) \ket{00\dots0} }^2\\
    &\quad = \omega(\psi)^4/2^{4+9},
\end{split}
\end{align}
\end{subequations}
where $S_{\sigma'}$ is the unitary permutation operator implementing the permutation
$\sigma'=(1)(2,3,5,4,7)(6)(8,9,11,10,13)(12)(14,15,17,16)(18)$,
and the CNOTs are $C_{s'} = C_{1,7}\otimes C_{2,5}\otimes C_{3,6}\otimes C_{4,13}\otimes C_{8,11}\otimes C_{9,12}\otimes C_{10,16}\otimes C_{14,17}\otimes C_{15,18}$.

\section{Implementations on quantum computer}
\label{sec:experiment}

\subsection{States}
\label{sec:experiment.states}

\begin{figure}
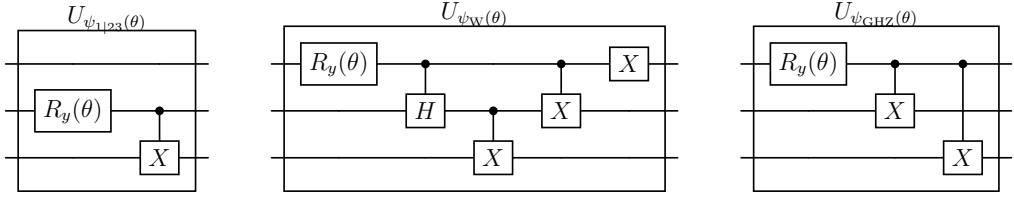

\centering
    \myquantikz{
     & \ghost{X}\gategroup[3,steps=2]{$U_{\psi_{1|23}(\theta)}$}
                &          & \\
     & \gate{R_y(\theta)} & \ctrl{1} & \\
     &          & \gate{X} &
    }
    \myquantikz{
     & \gate{R_y(\theta)}\gategroup[3,steps=5]{$U_{\psi_{\text{W}}(\theta)}$}
                & \ctrl{1} &          & \ctrl{1} & \gate{X} & \\
     &          & \gate{H} & \ctrl{1} & \gate{X} &          & \\
     &          &          & \gate{X} &          &          &
    }
    \myquantikz{
     & \gate{R_y(\theta)}\gategroup[3,steps=3]{$U_{\psi_{\text{GHZ}}(\theta)}$}
                & \ctrl{1} & \ctrl{2} & \\
     &          & \gate{X} &          & \\
     &\ghost{X} &          & \gate{X} &
    }
\caption{Quantum circuits implementing the preparations of the one-parameter families of three-qubit states~\eqref{eq:psitheta}.}
\label{fig:Us}
\end{figure}

The general form~\eqref{eq:Upsi222} of the preparation unitary $U_\psi$ is useful for theoretical calculations, but cannot directly be used in practice in quantum algorithms.
Implementing $U_\psi$ by decomposing it in terms of the usual quantum gates available in quantum computers
might be a challenging task for a particular state vector $\ket{\psi}$.
Here we consider the one-parameter families of state vectors
\begin{subequations}
\label{eq:psitheta}
\begin{align}
    \label{eq:psitheta.1v23}
    \ket{\psi_{1|23}(\theta)}       &:= \cos(\theta/2)\ket{000}+\sin(\theta/2)\ket{011},\\
    \label{eq:psitheta.W}
    \ket{\psi_{\text{W}}(\theta)}   &:= \cos(\theta/2)\ket{100}+\sin(\theta/2)\bigl(\ket{010}+\ket{001}\bigr)/\sqrt{2},\\
    \label{eq:psitheta.GHZ}
    \ket{\psi_{\text{GHZ}}(\theta)} &:= \cos(\theta/2)\ket{000}+\sin(\theta/2)\ket{111}.
\end{align}
\end{subequations}
The usual representing elements of the classes are also covered,
\begin{subequations}
\label{eq:psirepr}
\begin{align}
    \ket{\psi_{1|23}} &:= \ket{\psi_{1|23}(\theta_{1|23})} = \bigl(\ket{000}+\ket{011}\bigr)/\sqrt{2},& 
    \theta_{1|23} &= 2\arccos(1/\sqrt{2}),\\
    \ket{\psi_\text{W}} &:= \ket{\psi_{\text{W}}(\theta_\text{W})} = \bigl(\ket{001}+\ket{010}+\ket{100}\bigr)/\sqrt{3},&
    \theta_\text{W} &= 2\arccos(1/\sqrt{3}),\\
    \ket{\psi_\text{GHZ}} &:= \ket{\psi_{\text{GHZ}}(\theta_\text{GHZ})} = \bigl(\ket{000}+\ket{111}\bigr)/\sqrt{2},&
    \theta_\text{GHZ} &= 2\arccos(1/\sqrt{2}).
\end{align}    
\end{subequations}
For the one-parameter families of state vectors~\eqref{eq:psitheta},
the decomposition of $U_\psi$ in terms of the usual unitary gates available in quantum computers is not difficult.
We give a particular decomposition in Figure~\ref{fig:Us}.
Note that these state vectors are given by real coefficients,
and our implementation uses operators having real matrices in the computational basis,%
\footnote{Here $R_y(\theta)=e^{-i(\theta/2)Y}=\cos(\theta/2)I-i\sin(\theta/2)Y$.}
so $U_\psi$ can be used instead of $\overline{U}_\psi$, and $U_\psi^\dagger$ instead of $U_\psi^\T$.
$U_\psi^\dagger$, on the other hand, is just the reversed circuit with the parameter $-\theta$ instead of $\theta$ for these decompositions.

For the illustrations
we consider the LU-invariants $c_a^2$ \eqref{eq:222invs.ca}, $\omega^2$ \eqref{eq:222invs.omega} and $\tau^2$  \eqref{eq:222invs.tau},
as these are sufficient for the SLOCC classification,
and the square roots of these are important entanglement measures.
Using the explicit forms in Appendix~\ref{sec:explicit.gen},
straightforward calculations lead to the $\theta$-dependent values of these invariants
for the one-parameter families of state vectors~\eqref{eq:psitheta}.
These are shown in Table~\ref{tab:invspsitheta}.
Using the SLOCC classification in Table~\ref{tab:222classes}, formulated by the vanishing of the invariants,
we have that 
$\ket{\psi_{1|23}(\theta)}$ is in class $1|2|3$ (fully separable) for $\theta\in\pi\mathbb{N}$,
otherwise is in class $1|23$;
$\ket{\psi_{\text{W}}(\theta)}$ is in class $1|2|3$ for $\theta\in2\pi\mathbb{N}$,
it is in class $1|23$ for $\theta\in2\pi(\mathbb{N}+1/2)$,
otherwise it is in class W;
and $\ket{\psi_{\text{GHZ}}(\theta)}$ is in class $1|2|3$ for $\theta\in\pi\mathbb{N}$,
otherwise it is in class GHZ.

\begin{table}
\caption{The exact values of the invariants
for the state vectors~\eqref{eq:psitheta}.}
\label{tab:invspsitheta}
\setlength{\tabcolsep}{8pt}
\renewcommand{\arraystretch}{1.25}
\begin{tabular}{c||ccc|c|c}
    $\ket{\psi}$ &
    $c^2_1(\psi)$ & $c^2_2(\psi)$ & $c^2_3(\psi)$ &
    $\omega^2(\psi)$ &
    $\tau^2(\psi)$  \\
    \hline
    \hline
    $\ket{\psi_{1|23}(\theta)}$       & $0$ & $\sin^2(\theta)$ & $\sin^2(\theta)$     & $0$     & $0$ \\
    \hline
    $\ket{\psi_{\text{W}}(\theta)}$   & \parbox{2cm}{$\sin^2(\theta)/2+\sin^4(\theta/2)$} & \parbox{2cm}{$\sin^2(\theta)/2+\sin^4(\theta/2)$}& $\sin^2(\theta)$     & $\sin^2(\theta/2)\sin^2(\theta)$     & $0$ \\
    $\ket{\psi_{\text{GHZ}}(\theta)}$ & $\sin^2(\theta)$ & $\sin^2(\theta)$ & $\sin^2(\theta)$ & $\sin^2(\theta)$ & $\sin^4(\theta)$
\end{tabular}
\end{table}

\subsection{Results}
\label{sec:experiment.results}

In our experiments we used \texttt{ibmq\_pittsburgh}, which is one of the IBM Quantum Heron r3 processors~\cite{IBMQuantum-2026}.
The Qiskit worksheets~\cite{qiskit2024} containing the implementation of the invariants and running the experiments are attached as a supplementary.

\textit{First}, we consider the representing states~\eqref{eq:psirepr} only,
which are maximally entangled inside the corresponding SLOCC classes~\cite{Szalay-2025}.
We applied the smaller circuit 
for the implementation of the LU-invariants $c_a^2$, $\omega^2$ and $\tau^2$,
given in~\eqref{eq:p222.c1},~\eqref{eq:p222.omega} and~\eqref{eq:p222.tau}.
The invariants were measured for $10$ different instances of states,
chosen randomly from the LU-orbits of the representing states~\eqref{eq:psirepr}.
(We applied local unitary rotations by random angles.)
The mean value and the variance of the invariants are shown in Table~\ref{tab:invspsithetaexp}.
Here we have some small but significant error due to the noise.

\begin{table}
\caption{The measured values of the LU-invariants 
for $10$ instances of states in the LU-orbits of the state vectors~\eqref{eq:psirepr}.}
\label{tab:invspsithetaexp}
\begin{tabular}{c||ccc|c|c}
    $\ket{\psi}$ &
    $c^2_1(\psi)$ & $c^2_2(\psi)$ & $c^2_3(\psi)$ &
    $\omega^2(\psi)$ &
    $\tau^2(\psi)$  \\
    \hline
    \hline
    $\ket{\psi_{1|23}}$ & $0.114\cdot 10^{-2}$ & $0.965$ & $0.971$ & $0.113\cdot 10^{-1}$ & $0.004\cdot{10}^{-2}$ \\
    var: & $0.2 \cdot 10^{-4}$ & $1.416 \cdot 10^{-4}$ & $8.380 \cdot 10^{-5}$ & $2 \cdot 10^{-5}$ & $7.111 \cdot 10^{-7}$ \\
    exact: & $0$ & $1$ & $1$ & $0$ & $0$ \\
    \hline
    $\ket{\psi_\text{W}}$ & $0.855$ & $0.842$ & $0.842$ & $0.545$ & $0.007$ \\
    var: & $2.983 \cdot 10^{-5}$ & $2.130 \cdot 10^{-4}$ & $1.317 \cdot 10^{-5}$ & $2.044 \cdot 10^{-4}$ & $5.493 \cdot 10^{-7}$ \\
    exact: & $8/9 = 0.889$ & $8/9 = 0.889$ & $8/9 = 0.889$ & $16/27 = 0.593$ & $0$ \\
    \hline
    $\ket{\psi_\text{GHZ}}$ & $0.970$ & $0.959$ & $0.956$ & $0.961$ & $0.948$ \\
    var: & $7.500 \cdot 10^{-5}$ & $5.116 \cdot 10^{-5}$ & $7.741 \cdot 10^{-5}$ & $1.747 \cdot 10^{-4}$ & $8.575 \cdot 10^{-4}$ \\
    exact: & $1$ & $1$ & $1$ & $1$ & $1$
\end{tabular}
\end{table}

\begin{figure}
    \centering
    \includegraphics{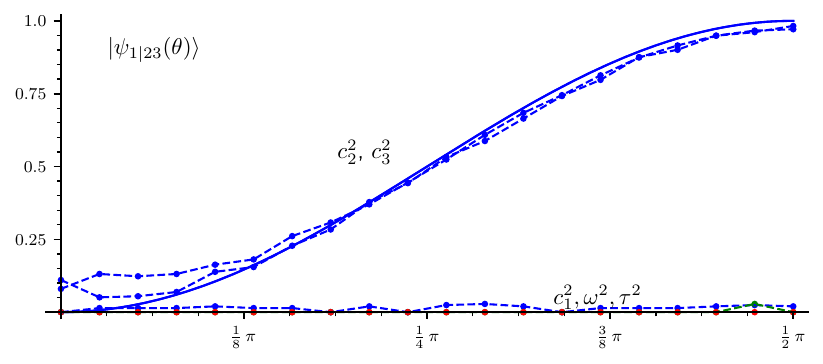}
    \includegraphics{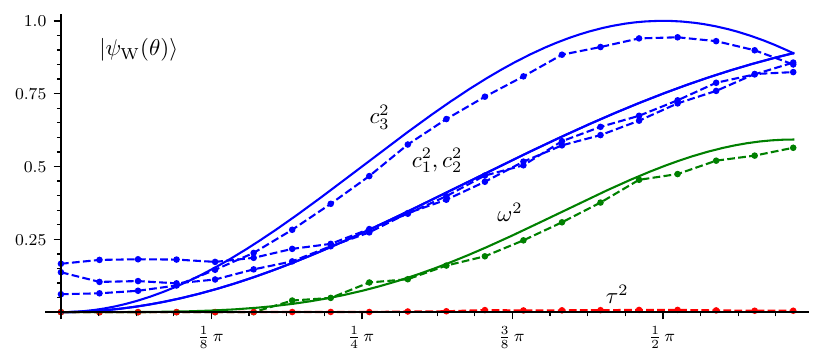}
    \includegraphics{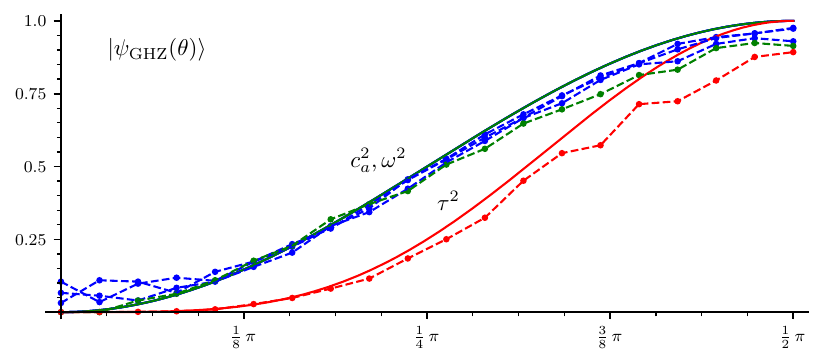}
    \caption{Values of the invariants $c_a^2$ (blue), $\omega^2$ (green) and $\tau^2$ (red)
    for the three one-parameter states~\eqref{eq:psitheta}.
    The exact values of the invariants (Table~\ref{tab:invspsitheta}) are shown by continuous curves,
    experimental values are shown by dots, connected by dashed lines to help to follow the data.}
    \label{fig:psitheta}
\end{figure}

\textit{Second}, we consider the one-parameter families of states~\eqref{eq:psitheta}.
We applied the smaller circuit 
for the implementation of the LU-invariants $c_a^2$, $\omega^2$ and $\tau^2$,
given in~\eqref{eq:p222.c1},~\eqref{eq:p222.omega} and~\eqref{eq:p222.tau}.
The LU-invariants measured for the $\theta$-dependent states are shown in Figure~\ref{fig:psitheta},
together with the plots of the exact values (Table~\ref{tab:invspsitheta}).
Here we have a small undermeasuring in general, but an overmeasuring for the small values of $c_a^2$,
and $\omega^2$.

\begin{figure}
    \centering
    \includegraphics{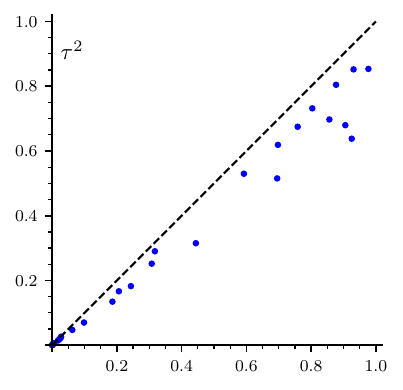}
    \includegraphics{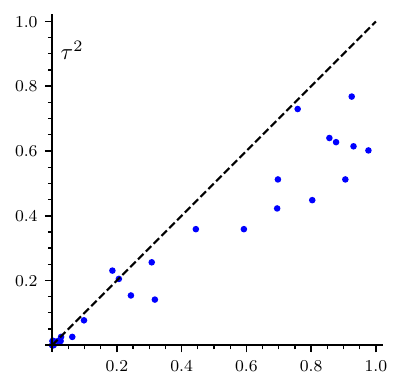}
    \caption{Measured values of the invariant $\tau^2$ against the calculated values,
    using the smaller~\eqref{eq:p222.tau} and the larger~\eqref{eq:p222B.tau} circuits.}
    \label{fig:randxxtausq}
\end{figure}

\textit{Third}, we consider the one-parameter family~\eqref{eq:psitheta.GHZ} of GHZ state vectors,
we used $3$ instances with randomly chosen $\theta$ values, followed by local $y$-rotations with randomly chosen angles.
To measure $\tau^2$, we applied both the smaller and the (second variant of the) larger circuit for the implementation of $\tau^2$, given in~\eqref{eq:p222.tau} and \eqref{eq:p222B.tau}, using $6$ and $12$ qubits, respectively.
The measured values against the calculated values of $\tau^2$ are shown in Figure~\eqref{fig:randxxtausq}.
Here we can see the better performance of the smaller circuit.

\section{Summary and remarks}
\label{sec:summary}

We have shown and illustrated \emph{two general methods} for the implementation
of the \emph{direct measurement} of important nonnegative (real-valued) LU-invariants of two and three qubits on quantum computers.
These translate the index contractions in the definitions of the LU-invariants to the language of quantum circuits,
then the probability of a particular outcome of the measurement of the qubit registers gives the absolute value squared of the index contraction scheme.
Our first method is more feasible than the second one and the earlier ad-hoc methods,
as using half as many qubit registers and measurements. 

Note that our first method is related to the \emph{interferometric method}~\cite{Horodecki-2002,Ekert-2002,Leifer-2004},
which, contrary to our methods, is capable to measure also the phase of an LU-invariant.
(For a complex invariant, see for example Grassl's invariant~\cite{Acin-2001}, needed to fully distinguish the LU orbits of three qubit state vectors.)
Note that the controlled permutation gate used in the interferometric setting is much more costly.
The straightforward generalization of our standalone permutation gate, using CNOT gates, to the interferometric setting would use three-qubit CCNOT (Toffoli) gates. 
It is known that five two-qubit gates are necessary and sufficient to implement CCNOT~\cite{Yu-2013}.

We have also performed some experiments on quantum computers using the IBM Quantum Platform.
We experienced small but significant error due to the noise or other imperfections inevitable in these devices.

Because of the imperfections, a fully accurate detection of SLOCC classes cannot be achieved by the present NISQ devices,
and it is easy to understand, why.
The class $1|2|3$ is of zero measure in the closures of the classes $a|bc$, which are of zero measures in the closure of the class W, which is of zero measure in the whole state space.
These closures are algebraic varieties given by the zero-loci of the corresponding invariants (see Table~\ref{tab:222classes} and \cite{holweck2012geometric}), giving a stratification of the space of state vectors (given by $n(\psi)=1$).
We have two roles of the errors here.
On the one hand, any preparation-error or gate-error leads to a state in class GHZ, as the other classes are of zero measure.
On the other hand, measurement errors may lead to false zeroes, not detecting weak entanglement.
The topology of the machine also has incidence because of the contraction operators. As we have seen, contraction of indices connects different qubits of the copies of our states. In this work we use the circuit optimization provided by Qiskit for our online experiments, but more could probably be done if one considers the physical connectivity of the machine.
Note, however, that despite these inevitable errors, it is still meaningful to consider the LU-invariants $c_a$, $\omega$ and $\tau$ as characterizing the SLOCC properties:
$c_a(\psi)$ tells us ``how much'' $\ket{\psi}$ is $a|bc$-entangled,
$\omega(\psi)$ tells us ``how much'' $\ket{\psi}$ is W-entangled,
$\tau(\psi)$ tells us ``how much'' $\ket{\psi}$ is GHZ-entangled.
``How much'' is meaningful now, it is given in terms of the different kinds of entanglement,
as all of these quantities are entanglement measures~\cite{Szalay-2025}.
A nonzero but small value means that the given multipartite entanglement property of the state is weak (allowing zero due to errors).

Although we considered two- and three-qubit invariants only, we emphasize that our methods are completely general,
working for arbitrary $n$-qubit invariants, if those are written in terms of local index contractions by $\epsilon$ and $\delta$.
Moreover, although only qubits are available in present days quantum devices, the use of our methods are not even restricted to qubit systems.
This is because, although the LU-invariants~\cite{Szalay-2025} considered here were given by LSL-covariants~\cite{Luque-2007},
where the $\epsilon$-contractions could be translated to the language of quantum circuits directly,
the LU-invariants in general can be formulated by local index contractions by $\delta$, working for all local dimensions.
In general, the $\delta$-form of the LU-invariants given by $\epsilon$-contractions in qubit systems 
can be obtained by the identity $\epsilon_{ii'}\epsilon_{kk'} = \delta_{ik}\delta_{i'k'} - \delta_{ik'}\delta_{i'k}$,
also for the LU-invariants considered here~\cite{Szalay-2025}.
Then the LU-invariant in question may be the linear combination of more elementary invariants given by $\delta$-contractions.
Note also that in the concrete implementation we applied the circuits for quantum states which were easy to implement,
that is, the oracle $U_\psi$ was easy to decompose into elementary operations available in quantum computers.
However, the circuits can be applied to any quantum state $\ket{\psi}$ after working out the implementation of a particular $U_\psi$ preparing it.

Comparing to the \emph{full tomography} of the three-qubit quantum state, which would require $3^3=27$ distinguished types of three-qubit measurements before evaluating those invariants from their algebraic expression, our methods require only one measurement of either a $3k/4$ or a $3k/2$ qubit circuit, where $k$ is the degree of the invariant of interest.
For the general case of $n$-qubit invariants, the \emph{full tomography} would require $3^n$ three-qubit measurements,
while our methods only require one measurement of either a $nk/4$ or a $nk/2$ qubit circuit.
In other words, while the number of measurements scales exponentially with the number of qubits in full tomography, our methods need only one measurement, while the number of qubits required still scale only linearly with the degree of the invariant we want to evaluate.

Interestingly, the idea of using algebraic geometry and invariant theory to measure entanglement and describe entanglement classes has been considered with different flavors over the past 25 years \cite{miyake2003classification,holweck2014entanglement, holweck2017entanglement,sanz2017entanglement,gharahi2024classifying}. The access to quantum computers allows us to turn those abstract and  almost two century old concepts, like Cayley's hyperdeterminant, into physical experiences. The evaluation of those invariants can now be used to benchmark current online quantum computers.

\section*{Acknowledgments}
We thank Péter Lévay and Péter Vrana for useful discussions.
Sz.Sz.~is partially supported by the Hungarian National Research, Development and Innovation Office within the grant K-134983, within the ``Frontline'' Research Excellence Programme KKP-133827, and within the Quantum Information National Laboratory of Hungary.
F.H.~is partially supported by the TACTICQ project (ANR-17-EURE-0002). F.H. acknowledges the support of the Erasmus staff mobility program, which allowed him to visit Budapest University of Technology and Economics on February 2025.
Sz.Sz.~happily acknowledges the support of the wonderful Bach performances of Marta Czech and Sir András Schiff.

We acknowledge the use of IBM Quantum Credits for this work.
The views expressed are those of the authors, and do not reflect the official policy or position of IBM or the IBM Quantum team.

No cats were harmed during the making of this work. Two of them were actually treated pretty well.

\appendix
\section{Error estimation}
\label{sec:error}

In our methods presented in the main text, 
the circuits obtaining an invariant of degree $k$ use $k/2$ copies of the state,
that is, $k/2$ instances of the oracle $U_\psi$.
The different instances may produce states with different error terms,
$\ket{\psi_r}:=\ket{\psi}+\ket{\xi_r}$ instead of $\ket{\psi}$,
where $\norm{\xi_r} = \norm{\ket{\psi_r}-\ket{\psi}}\leq \varepsilon$ is small.

Now consider a circuit implementing an invariant $f(\psi)$ in general.
We write it also denoting the dependency on all the instances explicitly, $f(\psi_1,\psi_2,\dots,\psi_{k/2})$.
Note that this is linear (or conjugate-linear) in every variable,
and the original value without error is $f(\psi) = f(\psi,\psi,\dots,\psi)$, being of degree $k/2$.
Then we have the estimation
\begin{equation}
\begin{split}
    &f(\psi_1,\psi_2,\dots,\psi_{k/2})
    = f(\psi+\xi_1,\psi+\xi_2,\dots,\psi+\xi_{k/2}) \\
    &\qquad= f(\psi,\psi,\dots,\psi) \\
    &\qquad\qquad+ f(\xi_1,\psi,\dots,\psi) + f(\psi,\xi_2,\dots,\psi) + \dots + f(\psi,\psi,\dots,\xi_{k/2})\\
    &\qquad\qquad+ \text{higher order terms}\\
    &\qquad= f(\psi) + \sum_{r=1}^{k/2} \braket{f_r(\psi)\vert\xi_r}^{(*)} + \text{higher order terms}.
\end{split}
\end{equation}
Here $\ket{f_r(\psi)}$ is the vector containing the coefficients multiplying those of the error term $\ket{\xi_r}$.
When $f$ depends on the complex conjugate $\ket{\overline{\psi_r}}$, then we have the complex conjugate $\braket{f_r(\psi)\vert\xi_r}^*$ in the first order error term, that is denoted by $(*)$.
The first order error term can be bounded as
\begin{equation}
    \Bigabs{\sum_{r=1}^{k/2} \braket{f_r(\psi)\vert\xi_r}^{(*)}}
    \leq \sum_{r=1}^{k/2} \abs{\braket{f_r(\psi)\vert \xi_r}}
    \leq \sum_{r=1}^{k/2} \norm{f_r(\psi)} \norm{\xi_r}
\end{equation}
by the triangle inequality of $\abs{\cdot}$ and the Cauchy–Bunyakovsky–Schwarz inequality.
It will turn out that $\norm{f_r(\psi)}\leq\norm{\psi}^{k/2-1}$ for these circuits, then we have the $\varepsilon k/2$ upper bound on the first order error in general.

Calculating upper bounds on the $\norm{f_r(\psi)}$ weights of the error magnitudes $\norm{\xi_r}$ is straightforward.
We basically have to carefully track the index contractions in the formulas.
The best way to do this is to use the diagrammatic presentation of the invariants, which are just the circuits themselves.
Then the application of the Cauchy–Bunyakovsky–Schwarz inequality, the tensor-multiplicatiy and submultiplicativity of the Hilbert-Schmidt norm give the upper bounds.

The two-qubit case is simple.
For the norm $n^2$~\eqref{eq:22tens.n} we have $\norm{n^2_{r=1,2}(\psi)}=\norm{\psi}=1$.
For $q$~\eqref{eq:22tens.q} we have $\norm{q_{r=1,2}(\psi)}=\norm{\tilde{\psi}}=\norm{\psi}=1$,
where $\ket{\tilde{\psi}}$ is the spin-flip~\cite{Wootters-1998,Szalay-2025}.
The upper bound on the first order error is then $\norm{\xi_1}+\norm{\xi_2}\leq 2\varepsilon$ in both cases.

The three-qubit case is more difficult, although the bounds themselves turn out to be simple, thanks to the structure of the invariants.
For the norm $n^2$~\eqref{eq:222tens.n} we have $\norm{n^2_{r=1,2}(\psi)}=\norm{\psi}=1$ again.
For the further invariants, $\norm{f_r(\psi)}$ can be written in terms of 
the state $\rho:=\ket{\psi}\bra{\psi}$ and the reduced states $\rho_{bc}:=\Tr_a(\rho)$, $\rho_a:=\Tr_{bc}(\rho)$.
Note that $\norm{\rho_{bc}}=\norm{\rho_a}$ since $\rho$ is pure, and $1/2\leq\norm{\rho_a}\leq1$ for any one-qubit state $\rho_a$.
For $g_a^2$~\eqref{eq:222invs.ga} we have 
\begin{equation*}
\norm{g_{a,r=1,\dots,4}^2(\psi)} = \Tr((\rho_a\otimes\tilde{\rho}_{bc})\rho)^{1/2}
\leq \norm{\rho_a}^{1/2}\norm{\tilde{\rho}_{bc}}^{1/2}\norm{\rho}^{1/2}
= \norm{\rho_a}\norm{\psi} = \norm{\rho_a}.
\end{equation*}
For $c_a^2$~\eqref{eq:222invs.ga} we have
\begin{equation*}
\norm{c_{a,r=1,\dots,4}^2(\psi)} = \Tr((\tilde{\rho}_a\otimes\rho_{bc})\rho)^{1/2}
\leq \norm{\tilde{\rho_a}}^{1/2}\norm{\rho_{bc}}^{1/2}\norm{\rho}^{1/2}
= \norm{\rho_a}\norm{\psi} = \norm{\rho_a}.
\end{equation*}
For $\omega^2/4$~\eqref{eq:222invs.omega} we have 
\begin{align*}
\norm{\omega_{r=1,4}^2(\psi)}/4 &= \Tr((\rho_a\tilde{\rho}_a\rho_a\otimes\rho_{bc})\tilde{\rho})^{1/2}
\leq \norm{\rho_a\tilde{\rho}_a\rho_a}^{1/2}\norm{\rho_{bc}}^{1/2}\norm{\tilde{\rho}}^{1/2}\\
&\leq \norm{\rho_a}^{1/2}\norm{\tilde{\rho}_a}^{1/2}\norm{\rho_a}^{1/2}\norm{\rho_{bc}}^{1/2}\norm{\rho}^{1/2}
= \norm{\rho_a}^2\norm{\psi} = \norm{\rho_a}^2,\\
\norm{\omega_{r=2,5}^2(\psi)}/4 &= \Tr((\rho_a^2\otimes\rho_{bc}^2)\tilde{\rho})^{1/2}
\leq \norm{\rho_a}\norm{\rho_{bc}}\norm{\tilde{\rho}}^{1/2}
= \norm{\rho_a}^2\norm{\psi} = \norm{\rho_a}^2\\
\norm{\omega_{r=3,6}^2(\psi)}/4 &= \Tr((\rho_a\otimes\rho_{bc}\tilde{\rho}_{bc}\rho_{bc})\tilde{\rho})^{1/2}
\leq \norm{\rho_a}^{1/2}\norm{\rho_{bc}\tilde{\rho}_{bc}\rho_{bc}}^{1/2}\norm{\tilde{\rho}}^{1/2}\\
&\leq \norm{\rho_a}^{1/2}\norm{\rho_{bc}}^{1/2}\norm{\tilde{\rho}_{bc}}^{1/2}\norm{\rho_{bc}}^{1/2}\norm{\rho}^{1/2}
= \norm{\rho_a}^2\norm{\psi} = \norm{\rho_a}^2.
\end{align*}
For $q$~\eqref{eq:222tens.q} we have
\begin{equation*}
\norm{q_{r=1,\dots,4}(\psi)} = \Tr((\rho_a\otimes\rho_{bc})\tilde{\rho})^{1/2}
\leq \norm{\rho_a}^{1/2}\norm{\rho_{bc}}^{1/2}\norm{\rho}^{1/2}
= \norm{\rho_a}\norm{\psi} = \norm{\rho_a}.
\end{equation*}
The upper bound on the first order error is then
$\norm{\xi_1}+\norm{\xi_2}\leq 2\varepsilon$ for $n^2$; while for the other circuits we have state-dependent bounds
$\norm{\rho_a}\sum_{r=1}^4\norm{\xi_r}\leq 4\varepsilon\norm{\rho_a}\leq 4\varepsilon$ for $g_a^2$, $c_a^2$, $q$; and
$\norm{\rho_a}^2\sum_{r=1}^6\norm{\xi_r}\leq 6\varepsilon\norm{\rho_a}^2\leq 6\varepsilon$ for $\omega^2/4$.
Note that $\norm{\rho_a}^2=(1-c_a^2(\psi)/2)$, so we have stronger bounds when the $a|bc$-entanglement of the state $\ket{\psi}$ is stronger.

\section{Explicit forms}
\label{sec:explicit}

\subsection{In general}
\label{sec:explicit.gen}

Here we show the explicit forms of the three-qubit invariants and covariants~\eqref{eq:222tens} appearing in the main text
for the general state $\ket{\psi}=\sum_{ijk} \psi^{ijk}\ket{ijk}$.

The quadratic covariants (symmetric tensors) given in Eqs.~\eqref{eq:222tens.gam1}-\eqref{eq:222tens.gam3},
$\ket{\gamma_a(\psi)}=\sum_{ii'}\gamma_1(\psi)^{ii'}\ket{ii'}\in \mathcal{H}_a\otimes\mathcal{H}_a$,
are of the form
\begin{subequations}
\begin{align}
    \gamma_1(\psi)= \begin{pmatrix}
        2\psi^{000}\psi^{011} -2\psi^{001}\psi^{010} \\
         \psi^{000}\psi^{111} - \psi^{001}\psi^{110} - \psi^{010}\psi^{101} + \psi^{011}\psi^{100} \\
         \psi^{100}\psi^{011} - \psi^{101}\psi^{010} - \psi^{110}\psi^{001} + \psi^{111}\psi^{000} \\
        2\psi^{100}\psi^{111} -2\psi^{101}\psi^{110}
    \end{pmatrix},\\
    \gamma_2(\psi)= \begin{pmatrix}
        2\psi^{000}\psi^{101} -2\psi^{001}\psi^{100} \\
         \psi^{000}\psi^{111} - \psi^{001}\psi^{110} - \psi^{100}\psi^{011} + \psi^{101}\psi^{010} \\
         \psi^{010}\psi^{101} - \psi^{011}\psi^{100} - \psi^{110}\psi^{001} + \psi^{111}\psi^{000} \\
        2\psi^{010}\psi^{111} -2\psi^{011}\psi^{110}
    \end{pmatrix},\\
    \gamma_3(\psi)= \begin{pmatrix}
        2\psi^{000}\psi^{110} -2\psi^{010}\psi^{100} \\
         \psi^{000}\psi^{111} - \psi^{010}\psi^{101} - \psi^{100}\psi^{011} + \psi^{110}\psi^{001} \\
         \psi^{001}\psi^{110} - \psi^{011}\psi^{100} - \psi^{101}\psi^{010} + \psi^{111}\psi^{000} \\
        2\psi^{001}\psi^{111} -2\psi^{011}\psi^{101}
    \end{pmatrix}.
\end{align}
The cubic covariant (tensor) given in Eq.~\eqref{eq:222tens.T},
$\ket{T(\psi)}=\sum_{ii'}T(\psi)^{ijk}\ket{ijk}\in\mathcal{H}_1\otimes\mathcal{H}_2\otimes\mathcal{H}_3$,
is of the form
\begin{equation}
    T(\psi)= \begin{pmatrix}
        + \psi^{000}( \psi^{000}\psi^{111} - \psi^{100}\psi^{011} - \psi^{010}\psi^{101} - \psi^{001}\psi^{110} ) + 2\psi^{100}\psi^{010}\psi^{001} \\
        - \psi^{001}( \psi^{001}\psi^{110} - \psi^{101}\psi^{010} - \psi^{011}\psi^{100} - \psi^{000}\psi^{111} ) - 2\psi^{101}\psi^{011}\psi^{000} \\
        - \psi^{010}( \psi^{010}\psi^{101} - \psi^{110}\psi^{001} - \psi^{000}\psi^{111} - \psi^{011}\psi^{100} ) - 2\psi^{110}\psi^{000}\psi^{011} \\
        + \psi^{011}( \psi^{011}\psi^{100} - \psi^{111}\psi^{000} - \psi^{001}\psi^{110} - \psi^{010}\psi^{101} ) + 2\psi^{111}\psi^{001}\psi^{010} \\
        - \psi^{100}( \psi^{100}\psi^{011} - \psi^{000}\psi^{111} - \psi^{110}\psi^{001} - \psi^{101}\psi^{010} ) - 2\psi^{000}\psi^{110}\psi^{101} \\
        + \psi^{101}( \psi^{101}\psi^{010} - \psi^{001}\psi^{110} - \psi^{111}\psi^{000} - \psi^{100}\psi^{011} ) + 2\psi^{001}\psi^{111}\psi^{100} \\
        + \psi^{110}( \psi^{110}\psi^{001} - \psi^{010}\psi^{101} - \psi^{100}\psi^{011} - \psi^{111}\psi^{000} ) + 2\psi^{010}\psi^{100}\psi^{111} \\
        - \psi^{111}( \psi^{111}\psi^{000} - \psi^{011}\psi^{100} - \psi^{101}\psi^{010} - \psi^{110}\psi^{001} ) - 2\psi^{011}\psi^{101}\psi^{110}
    \end{pmatrix}.
\end{equation}
Note that it can also be written as
\begin{equation}
    (-1)^{i+j+k} T(\psi)^{ijk} =
    \psi^{ijk}\bigl(
          \psi^{ijk}\psi^{\nn{i}\nn{j}\nn{k}} 
        - \psi^{\nn{i}jk}\psi^{i\nn{j}\nn{k}}
        - \psi^{i\nn{j}k}\psi^{\nn{i}j\nn{k}}
        - \psi^{ij\nn{k}}\psi^{\nn{i}\nn{j}k}
    \bigr)
    + 2\psi^{\nn{i}jk}\psi^{i\nn{j}k}\psi^{ij\nn{k}},
\end{equation}
where the negation of the indices is $\nn{i}=1-i$.
The quartic invariant (scalar) given in Eq.~\eqref{eq:222tens.T},
$q(\psi)\in\field{C}$,
is of the form
\begin{equation}
\begin{split}
    q(\psi) = \hspace{9.8cm}&\\
    - 2\bigl(
        \psi^{000}\psi^{111}\psi^{000}\psi^{111}
      + \psi^{011}\psi^{100}\psi^{011}\psi^{100}
      + \psi^{101}\psi^{010}\psi^{101}\psi^{010}
     &+ \psi^{110}\psi^{001}\psi^{110}\psi^{001}
    \bigr)\\
    + 4\bigl(
        \psi^{000}\psi^{111}\psi^{011}\psi^{100}
      + \psi^{000}\psi^{111}\psi^{101}\psi^{010}
     &+ \psi^{000}\psi^{111}\psi^{110}\psi^{001}\\
      + \psi^{011}\psi^{100}\psi^{101}\psi^{010}
     &+ \psi^{011}\psi^{100}\psi^{110}\psi^{001}\\
     &+ \psi^{101}\psi^{010}\psi^{110}\psi^{001}
    \bigr)\\
    - 8\bigl(
        \psi^{000}\psi^{011}\psi^{101}\psi^{110}
     &+ \psi^{111}\psi^{100}\psi^{010}\psi^{001}
    \bigr).
\end{split}
\end{equation}
\end{subequations}

\subsection{For the LU-canonical form}
\label{sec:explicit.canon}

Here we show the explicit form of the three-qubit invariants and covariants~\eqref{eq:222tens}
and~\eqref{eq:222invs} appearing in the main text
for the LU-canonical form of three-qubit vectors~\cite{Acin-2000,Acin-2001}
\begin{equation}
\label{eq:canonical}
    \ket{\psi_\text{C}} := \lambda_0\ket{000}
    + e^{i\theta}\lambda_1\ket{100} + \lambda_2\ket{101}
               + \lambda_3\ket{110} + \lambda_4\ket{111},
\end{equation}
where $0\leq\lambda_i\in\field{R}$, $0\leq\theta\leq\pi$,
and $\norm{\psi_\text{C}}^2 = \sum_{k=0}^4\lambda_k^2$ is the norm.
Sometimes it is convenient to use the $\eta_k:=\lambda_k^2$ squares of the amplitudes,
and $\delta:=\lambda_1\lambda_4e^{i\theta} -\lambda_2\lambda_3$.

\begin{subequations}
The quadratic covariants (symmetric tensors) given in Eqs.~\eqref{eq:222tens.gam1}-\eqref{eq:222tens.gam3},
$\ket{\gamma_a(\psi_\text{C})}=\sum_{ii'}\gamma_1(\psi_\text{C})^{ii'}\ket{ii'}\in \mathcal{H}_a\otimes\mathcal{H}_a$,
are of the form
\begin{equation}
    \gamma_1(\psi_\text{C})= \begin{pmatrix}
        0 \\
        \lambda_0 \lambda_4 \\
        \lambda_0 \lambda_4 \\
        2\delta
    \end{pmatrix},\quad
    \gamma_2(\psi_\text{C})= \begin{pmatrix}
        2\lambda_0 \lambda_2 \\
        \lambda_0 \lambda_4 \\
        \lambda_0 \lambda_4 \\
        0
    \end{pmatrix},\quad
    \gamma_3(\psi_\text{C})= \begin{pmatrix}
        2\lambda_0 \lambda_3 \\
        \lambda_0 \lambda_4 \\
        \lambda_0 \lambda_4 \\
        0
    \end{pmatrix},
\end{equation}
The cubic covariant (tensor) given in Eq.~\eqref{eq:222tens.T},
$\ket{T(\psi_\text{C})}=\sum_{ii'}T(\psi_\text{C})^{ijk}\ket{ijk}\in\mathcal{H}_1\otimes\mathcal{H}_2\otimes\mathcal{H}_3$,
is of the form
\begin{equation}
    T(\psi_\text{C})= \begin{pmatrix}
        \lambda_0^2 \lambda_4 \\
        0 \\
        0 \\
        0 \\
        \lambda_0(\delta - \lambda_2 \lambda_3) \\
        - \lambda_0 \lambda_2 \lambda_4 \\
        - \lambda_0 \lambda_3 \lambda_4 \\
        - \lambda_0 \lambda_4^2
    \end{pmatrix}.
\end{equation}
The quartic invariant (scalar) given in Eq.~\eqref{eq:222tens.T},
$q(\psi_\text{C})\in\field{C}$,
is of the form
\begin{equation}
    q(\psi_\text{C}) = -2\lambda_0^2\lambda_4^2 e^{2 i\varphi}.
\end{equation}
\end{subequations}

\begin{subequations}
The LU-invariants given in Eqs.~\eqref{eq:222invs} are then of the form
\begin{align}
    y^2(\psi_\text{C})      &= 8(\eta_0\eta_2+\eta_0\eta_3+\abs{\delta}^2)/3+4\eta_0\eta_4, \\
    c_1^2(\psi_\text{C})    &= 4\eta_0(\eta_2+\eta_3+\eta_4), \\
    c_2^2(\psi_\text{C})    &= 4\eta_0(\eta_3+\eta_4) + 4\abs{\delta}^2, \\
    c_3^2(\psi_\text{C})    &= 4\eta_0(\eta_2+\eta_4) + 4\abs{\delta}^2, \\
    g_1^2(\psi_\text{C})    &= 2\eta_0\eta_4+4\abs{\delta}^2, \\
    g_2^2(\psi_\text{C})    &= 2\eta_0\eta_4+4\eta_0\eta_2, \\
    g_3^2(\psi_\text{C})    &= 2\eta_0\eta_4+4\eta_0\eta_3, \\
    \omega^2(\psi_\text{C}) &= 4\eta_0\eta_4 \norm{\psi_\text{C}}^2 - 16\eta_0\sqrt{\eta_2\eta_3}\Re(\delta), \\
    \tau^2(\psi_\text{C})   &= 16\eta_0^2\eta_4^2.
\end{align}
\end{subequations}

\bibliographystyle{unsrt}
\bibliography{references}{}

\end{document}